\documentclass[a4paper]{article}

\usepackage{jheppub}
\author{Marc-Antoine Fiset,}
\author{Matthias R.\ Gaberdiel,}
\author{Kiarash Naderi,}
\author{Vit Sriprachyakul}
\affiliation{Institut f\"ur Theoretische Physik,
ETH Z\"urich,\\
Wolfgang-Pauli-Stra{\ss}e 27,
8093 Z\"urich, Switzerland}
\emailAdd{fisetmarcantoine@gmail.com}
\emailAdd{gaberdiel@itp.phys.ethz.ch}
\emailAdd{knaderi@phys.ethz.ch}
\emailAdd{vsriprachyak@phys.ethz.ch}

\usepackage{bm} 
\usepackage{xcolor}
\definecolor{green_maf}{RGB}{28, 166, 46}
\definecolor{blue_mrg}{RGB}{12, 143, 145}
\definecolor{detail}{RGB}{110,110,110}
\usepackage{amsmath} 
\usepackage{nccmath} 
\usepackage{amssymb} 
\usepackage{amsthm} 
\usepackage{mathtools} 
\usepackage[utf8]{inputenc} 
\usepackage{braket}
\usepackage{enumerate}
\usepackage[inline]{enumitem}
\usepackage{comment}
\usepackage{soul} 

\usepackage{tikz}

\usetikzlibrary{calc,decorations.markings}
\usetikzlibrary{decorations.pathmorphing}
\usetikzlibrary{shapes,backgrounds}
\usetikzlibrary{fadings}
\usetikzlibrary{cd}
\usetikzlibrary{decorations.pathreplacing,calligraphy}

\tikzset{
partial ellipse/.style args={#1:#2:#3}{
insert path={+ (#1:#3) arc (#1:#2:#3)}
}
}

\newif\ifdetails
\detailstrue

\newcommand{\dd}{{\rm d}} 

\def\be{\begin{equation}}
\def\ee{\end{equation}}

\begin{document}

\title{Perturbing the symmetric orbifold from the worldsheet}

\abstract{The symmetric orbifold of $\mathbb{T}^4$ is the analogue of free SYM in four dimensions, and its dual is described by a tensionless string propagating in ${\rm AdS}_3\times {\rm S}^3 \times \mathbb{T}^4$. In this paper we study the deformation of this exact AdS/CFT duality away from the free point. On the symmetric orbifold side this amounts to perturbing the theory by the exactly marginal operator from the $2$-cycle twisted sector. We identify the corresponding perturbation in the dual worldsheet description, and show that the anomalous conformal dimensions of a number of symmetric orbifold currents are correctly reproduced  from this worldsheet perspective. }

\maketitle

\section{Introduction}

Recently some progress has been made towards identifying the exact AdS string theory that is dual to the free CFT on the boundary of AdS. In particular, for the case of the ${\rm AdS}_3/{\rm CFT}_2$ duality, the free CFT is to be identified with the symmetric orbifold of $\mathbb{T}^4$, and it was shown in \cite{Gaberdiel:2018rqv,Eberhardt:2018ouy} that the dual string theory on ${\rm AdS}_3 \times {\rm S}^3 \times \mathbb{T}^4$ has one unit ($k=1$) of NS-NS flux and can be described by a solvable worldsheet theory. Various consistency checks of this proposal have been performed; in particular, the spectrum and the amplitudes of the two descriptions have been matched with one another \cite{Eberhardt:2018ouy,Eberhardt:2019ywk,Eberhardt:2020akk,Dei:2020zui,Knighton:2020kuh}, see also \cite{Bertle:2020sgd,Dei:2021xgh,Gaberdiel:2021njm,Dei:2021yom,Gaberdiel:2022als,Dei:2022pkr,Gaberdiel:2022oeu,Naderi:2022bus} for related work. There has also been a suggestion for how this description should generalise to the case of ${\rm AdS}_5/{\rm CFT}_4$, for which the (free) boundary CFT is (free) ${\cal N}=4$ SYM in four dimensions \cite{Gaberdiel:2021iil,Gaberdiel:2021jrv}. 

While it is important to decipher how the duality works in detail at the free point, it is obviously even more interesting to understand how things work once one switches on the coupling constant of the dual CFT. In the context of large $N$ ${\cal N}=4$ SYM in four dimensions, the relevant coupling is the 't~Hooft coupling $\lambda$, and much detailed information is known about the anomalous dimensions of the operators for non-trivial $\lambda$, see e.g.\ \cite{Beisert:2010jr} for a review. Similarly, for the case of ${\rm AdS}_3/{\rm CFT}_2$, the relevant coupling describes the deformation of the symmetric orbifold theory by an exactly marginal operator from the $2$-cycle twisted sector, and again much is known about the anomalous dimensions of various CFT operators, see e.g.\ \cite{Burrington:2012yq,Gaberdiel:2015uca,Burrington:2017jhh,Guo:2019ady,Guo:2020gxm,Apolo:2022fya}. This deformation must have a counterpart on the worldsheet, and in this paper we explain how it can be described from that perspective. 
\smallskip

The basic idea of the construction is relatively straightforward. Since the exactly marginal operator $\Phi$ that deforms the symmetric orbifold theory away from the orbifold point is an element of the single particle CFT Hilbert space, it must appear in the physical spectrum of the worldsheet theory, i.e.\ there is a physical worldsheet operator ${\bf \Phi}$ that corresponds to it. Once the corresponding state has been identified --- this will be done using the DDF operators of \cite{Naderi:2022bus} ---  one can calculate the worldsheet correlators with the insertion of these perturbing fields, and this allows us to imitate the dual CFT perturbation on the worldsheet. For some simple examples --- in particular, we focus on certain spin-$1$ currents --- we show that this worldsheet approach reproduces the correct anomalous dimensions as calculated in the symmetric orbifold in \cite{Gaberdiel:2015uca,Apolo:2022fya}. The main technical hurdle for this calculation is to fix the overall normalisation factors since they are not directly determined by the Ward identities that underly the worldsheet calculations of \cite{Eberhardt:2019ywk,Eberhardt:2020akk,Dei:2020zui,Knighton:2020kuh}. In this paper we have solved this problem by noting that the same normalisation factors appear also in other amplitudes, which allows us to determine them indirectly. 

This shows, in principle, how the perturbed CFT can be determined from the worldsheet. However, the method is somewhat indirect in that the worldsheet calculation only determines the symmetric orbifold correlators with the insertion of the perturbing fields, and the actual perturbation calculation is then effectively performed in the dual CFT. On the other hand, it must also be possible to directly deform the worldsheet theory, and in fact we can interpret our analysis in this language: 
the perturbation of the dual CFT is obtained by adding 
\be
\int d^2z \, \int d^2x \, \bigl( \tilde{G}^-_{-1} \bar{\tilde{G}}^{-}_{-1} {\bf \Phi}\bigr)(z,x) \ , 
\ee
to the worldsheet action, where $\tilde{G}^-_{-1}$ and $\bar{\tilde{G}}^{-}_{-1}$ refer to the left- and right-moving supercurrents that appear in the ${\cal N}=4$ hybrid description of the worldsheet theory \cite{Berkovits:1999im}. Note that since ${\bf \Phi}$ has worldsheet conformal dimension $L_0=\bar{L}_0=0$, the $\tilde{G}^-_{-1}$ and $\bar{\tilde{G}}^{-}_{-1}$ descendants have $L_0=\bar{L}_0=1$, and hence the integrand of the worldsheet $z$-integral is marginal (and the integral conformally invariant). The $x$-integral then implements the perturbation of the spacetime CFT, and it also preserves the spacetime conformal symmetry since the perturbing field has $(h,\bar{h})=(1,1)$ in spacetime. Finally, one notes that because of the localisation properties of the correlators, there is in effect only a single integral that survives since all correlators involve a delta function relating $x$ to some function of $z$. 
\bigskip

The paper is organised as follows. In Section~\ref{2}, we briefly recall the set-up of \cite{Gaberdiel:2015uca} for the description of the dual CFT, and how anomalous dimensions can be obtained from the general theory of conformal perturbations. In Section~\ref{sec:hyb-states}, we review the hybrid formalism of the worldsheet theory, and identify the physical states we are interested in. In Section~\ref{fix}, we compute explicitly the 4-point correlators on the worldsheet, and then use them to determine the spacetime anomalous dimensions in Section~\ref{5}. Our conclusions are contained in Section~\ref{con}. There are a number of appendices where various conventions and technical details are described.

\section{Perturbing the symmetric orbifold}\label{2}

This section reviews the results of \cite{Gaberdiel:2015uca} for the calculation of the anomalous dimensions in the symmetric orbifold theory that we will later reproduce using the worldsheet description. We will follow the conventions of Appendix~\ref{app:N=4ofT4}, see also \cite{David:2002wn,Pakman:2009zz,Lunin:2000yv} for further information.

\subsection{Marginal operators in the twisted sector $(w=2)$} \label{sec:marginal-deformations}

The marginal operators we will be using arise from the 2-cycle twisted sector of the symmetric orbifold theory (corresponding to permutation cycles of length $w=2$). Focusing first on the left-moving degrees of freedom, the ground states of this sector have conformal weight $h=\tfrac{1}{2}$ and they form two $\mathfrak{su}(2)_R$ singlets and one doublet $\ket{\Omega^{w=2}_{\pm}}$. 

Since the elements of the doublet have charge $q = \pm h = \pm\tfrac{1}{2}$ (with respect to $\mathcal{J}^3$ in eq.~\eqref{eq:RsymmetryDual} below), they are chiral primaries (for $+$) and anti-chiral primaries (for $-$), respectively. 
As such, they are annihilated by half of the supercharges,
\begin{equation}
G^\pm_{-\frac{1}{2}}\ket{\Omega^{w=2}_{\pm}}={\tilde{G}}^\pm_{-\frac{1}{2}}\ket{\Omega^{w=2}_{\pm}}=0 \ .
\end{equation}
The other descendants yield two independent uncharged states with $h=1$:
\begin{equation}
\ket{\Phi}= G^-_{-\frac{1}{2}}\ket{\Omega^{w=2}_{+}}
= -\tilde{G}^+_{-\frac{1}{2}}\ket{\Omega^{w=2}_{-}}\ ,
\qquad\qquad
\ket{\tilde{\Phi}}= {\tilde{G}}^-_{-\frac{1}{2}}\ket{\Omega^{w=2}_{+}}
= G^+_{-\frac{1}{2}}\ket{\Omega^{w=2}_{-}}\ .
\end{equation}
 In addition, there is the residual torus symmetry $\mathfrak{su}(2)_B$ that acts on the bosonic modes, and the four  supercurrents sit in the $({\bf 2},{\bf 2})$ with respect to $\mathfrak{su}(2)_R \oplus \mathfrak{su}(2)_B$. Since the ground states are uncharged with respect to $\mathfrak{su}(2)_B$, the two singlet states $\ket{\Phi}$ and $\ket{\tilde{\Phi}}$ transform as $({\bf 1},{\bf 2})$, i.e.\ they form a doublet with respect to $\mathfrak{su}(2)_B$. As a consequence, their $2$-point functions satisfy $\braket{\Phi(x)\Phi(y)}=\braket{\tilde\Phi(x)\tilde\Phi(y)}=0$, as well as $\braket{\Phi(x)\tilde\Phi(y)}=\tfrac{1}{(x-y)^2}$.

In the full theory, these $\mathfrak{su}(2)_R$ singlet states are tensored with similar states from the right-moving sector, yielding altogether four marginal operators with $(h,\bar{h})=(1,1)$. Supersymmetry guarantees that they remain marginal to all orders in conformal perturbation theory upon the deformation that they induce, i.e.\ that they are exactly marginal. Since they come from a twisted sector, they mix untwisted and twisted sectors, and hence break the symmetric orbifold structure. There is good evidence, see e.g.\ \cite{Gaberdiel:2015uca, Apolo:2022fya}, that only the $\mathcal{N}=(4,4)$ superconformal currents survive under this deformation, while the other higher spin currents (that are present at the orbifold point) gain an anomalous dimension.

As in \cite{Gaberdiel:2015uca}, we will consider the exactly marginal operator that is a singlet with respect to the global $\mathfrak{so}(4) = \mathfrak{su}(2)_R \oplus \mathfrak{su}(2)_B$ symmetry, where by `global' we mean the diagonal action on both left- and right-movers. In terms of our conventions above, the relevant perturbing field is therefore 
\begin{equation} \label{eq:marginal_general}
\Phi(x,\bar{x}) = \frac{i}{\sqrt{2}} \Bigl( \Phi(x)\otimes\bar{\tilde \Phi}(\bar{x})
-  \tilde{\Phi}(x)\otimes\bar\Phi(\bar{x}) \Bigr) \ ,
\end{equation}
where we have denoted the right-movers by a bar, and chosen the coefficients so that the $2$-point function is normalised as 
\begin{equation}
\braket{\Phi(x,\bar{x})\Phi(y,\bar{y})}=\frac{1}{(x-y)^2(\bar x-\bar y)^2}\ .
\end{equation}

\subsection{Spin 1 currents in the untwisted sector $(w=1)$}

Next we introduce the currents of the symmetric orbifold theory whose anomalous dimensions we will calculate.


Consider first one copy of the $\mathbb{T}^4$ CFT, and focus on the left-moving sector. At spin $s=1$, there are six fermion bilinears which generate two commuting $\mathfrak{su}(2)_1$ Kac-Moody algebras.
One is the $\mathcal{N}=4$ R-symmetry $\mathfrak{su}(2)$, which we take to be generated by
\begin{equation} \label{eq:RsymmetryDual}
\mathcal{J}^3 = \frac{1}{2}(\psi^1\bar{\psi}^1 + \psi^2\bar{\psi}^2) \ , \qquad
\mathcal{J}^+= \psi^1\psi^2 \ , \qquad
\mathcal{J}^-= -\bar{\psi}^1\bar{\psi}^2 \ ,
\end{equation}
while the other is $\mathcal{N}=4$ primary and generated by
\begin{equation} \label{eq:other-su2}
\mathcal{M}^3 = \frac{1}{2}(\psi^1\bar{\psi}^1 - \psi^2\bar{\psi}^2) \ , \qquad
\mathcal{M}^+= \psi^1\bar{\psi}^2 \ , \qquad
\mathcal{M}^-= -\bar{\psi}^1\psi^2 \ .
\end{equation}
The fermions here satisfy the OPEs given in Appendix~\ref{app:N=4ofT4}.

An operator $O$ in one copy of the $\mathbb{T}^4$ CFT can be mapped to an operator in the orbifold $(\mathbb{T}^4)^{\otimes N}/S_N$ via
\begin{equation} \label{eq:map_to_untwisted_op}
O \longmapsto \frac{1}{\sqrt{N}}\, \sum_{n=1}^{N} \bigl({\bf 1}^{\otimes (n-1)} \otimes O\otimes {\bf 1}^{\otimes (N-n)} \bigr)\equiv \frac{1}{\sqrt{N}}\,  
\sum_{n=1}^N  O^{(n)} \ ,
\end{equation}
where  $O^{(n)}$ acts as $O$ in the $n$\textsuperscript{th} copy of the $\mathbb{T}^4$ CFT and trivially elsewhere. The resulting operator is in the untwisted sector. We will usually abuse notations by calling the image of eq.~\eqref{eq:map_to_untwisted_op} also $O$.

The operators corresponding to $\mathcal{M}^{3}$ and $\mathcal{M}^{\pm}$ via this prescription are the operators whose anomalous dimensions we will calculate. They are higher spin currents at the orbifold point with $(h,\bar{h})=(1,0)$, and they form the bottom component of a $R^{(1)}$ $\mathcal{N} = 4$ supermultiplet in the extended $\mathcal{W}_\infty[0]$ algebra of \cite{Gaberdiel:2013vva}.

\subsection{Anomalous dimensions from conformal perturbation theory} \label{sec:conf_pert_theory}

With these preparations at hand we can now summarise how the anomalous dimensions can be calculated in conformal perturbation theory. Let $W^{i}(x)$ be conformal quasi-primaries of the unperturbed theory with $(h,\bar{h})=(s, 0)$, normalised so that
\begin{equation}\label{Wnormalisation}
\langle W^{i}(x) W^{j}(y)\rangle = \frac{\delta^{ij}}{(x-y)^{2s}} \ .
\end{equation}
The perturbed correlators can be obtained from 
\begin{equation} \label{eq:deformed_corr}
\langle W^{i}(x) W^{j}(y)\rangle_\Phi
\equiv
\frac{\langle e^{\delta S} W^{i}(x) W^{j}(y) \rangle}{\langle e^{\delta S}\rangle}  \ ,
\qquad
\delta S = g_\text{CFT}\int \dd^2 x' \ \Phi(x^{\prime},\bar{x}^{\prime}) \ ,
\end{equation}
where $\Phi(x^{\prime},\bar{x}^{\prime})$ is the exactly marginal perturbing operator. Expanding the right-hand side of the first equation in eq.~\eqref{eq:deformed_corr} for small deformation parameter $g_\text{CFT}$ yields, up to second order 
\begin{align}
\langle W^{i}(x) W^{j}(y)\rangle_\Phi
= &\, \langle W^{i}(x) W^{j}(y)\rangle  \nonumber \\
&+ g_\text{CFT}\int \dd^2 x_1
\big\langle \Phi(x_1, \bar{x}_1) W^{i}(x) W^{j}(y)\big\rangle \nonumber \\
&+\frac{1}{2}g^2_\text{CFT} \int\dd^2 x_1 \dd^2 x_2\Big(
\big\langle \Phi(x_1, \bar{x}_1) \Phi(x_2, \bar{x}_2)W^{i}(x) W^{j}(y)\big\rangle  \label{eq:def_corr1} \\
&\qquad\qquad\qquad\qquad\qquad  - \big\langle \Phi(x_1, \bar{x}_1) \Phi(x_2, \bar{x}_2)\big\rangle \ \big\langle W^{i}(x) W^{j}(y)\big\rangle \Big) + \cdots \ . \nonumber
\end{align}
The term linear in $g_\text{CFT}$ vanishes. Indeed, since the $W^{i}(x)$ are left-moving currents, the correlator is in fact a 1-point correlator with respect to the right-movers ($\bar{x}_1$). Alternatively, in the case we are interested in, the 3-point correlator involves only one field from a twisted sector (and two fields from the untwisted sector), and hence again vanishes by the usual orbifold selection rules.

The leading correction appears therefore at order $O(g^2_\text{CFT})$, and it is due to the double integral of the \emph{connected} 4-point correlator. Amongst other terms, there will be a logarithmic contribution, turning eq.~\eqref{eq:def_corr1} into something of the form
\begin{equation} \label{eq:def_corr_log}
\langle W^{i}(x) W^{j}(y)\rangle_\Phi
=\frac{1}{(x-y)^{2s}}\left(\delta^{ij}-4\gamma^{ij}\ln|x-y| + \cdots\right) \ .
\end{equation}
Diagonalising the symmetric matrix $\gamma=(\gamma^{ij})$, the right-hand side eq.~\eqref{eq:def_corr_log} takes the form
\begin{equation}
\frac{\delta^{ij}}{(x-y)^{2s}}
\left(1-4\lambda^i\ln|x-y|+\cdots\right) \ ,
\label{2.15}
\end{equation}
where $\lambda^i$ are the eigenvalues.
These are the leading terms in an expansion of
\begin{equation}
\delta^{ij}(x-y)^{-2(s+\lambda^i)}(\bar{x}-\bar{y})^{-2\lambda^i}
\end{equation}
for small $\lambda^i$, thus identifying $\lambda^i$ as the anomalous dimensions of interest. Summarising, the anomalous dimensions are extracted by diagonalising the matrix of numerical coefficients of the logarithmic term $\ln|x-y|$ (and dividing by the overall factor of $-4$). The size of this matrix depends on how many currents mix non-trivially upon deformation. To leading order, only fields of the same conformal dimension can mix.

\subsection{4-point functions in the symmetric orbifold} \label{corr}

Let us now specialise to our situation, where the marginal operator is given by eq.~\eqref{eq:marginal_general}. Since the $W^i$ fields are purely left-moving, the $4$-point function appearing in eq.~(\ref{eq:def_corr1}) factorises as 
\begin{equation}\label{master} 
\begin{aligned}
\big\langle \Phi(x_1, \bar{x}_1) \Phi(x_2, \bar{x}_2)W^{i}(x) W^{j}(y)\big\rangle  & = \,  
\frac{1}{2} \, \big\langle \Phi(x_1) \tilde{\Phi}(x_2)\, W^{i}(x) W^{j}(y)\big\rangle\cdot  \big\langle \bar{\tilde{\Phi}}(\bar{x}_1) \bar{\Phi}(\bar{x}_2) \big\rangle  + (1 \leftrightarrow 2)\\
& = \frac{1}{2} \, \big\langle \Phi(x_1) \tilde{\Phi}(x_2)\, W^{i}(x) W^{j}(y)\big\rangle\cdot \frac{1}{(\bar{x}_1 - \bar{x}_2)^2} 
+ (1 \leftrightarrow 2) \ , 
\end{aligned}
\end{equation}
where, because of the anti-holomorphic correlator, only the mixed terms survive. Thus we will only need to calculate the chiral correlator 
\be\label{deform1}
\big\langle \Phi(x_1) \tilde{\Phi}(x_2)W^{i}(x) W^{j}(y)\big\rangle \ .
\ee

In order to fix various normalisation constants later on in the worldsheet calculations, it will be convenient to not just calculate this $4$-point function, but also the one  where we replace the left-moving part of $\Phi$ and $\tilde{\Phi}$ by the ground states in the $2$-cycle twisted sector $\Omega^{w=2}_{\pm}$,
\begin{equation}
\big\langle \Omega^{w=2}_+(x_1) \Omega^{w=2}_-(x_2)W^{i}(x) W^{j}(y)\big\rangle \ , 
\label{2.16}
\end{equation}
where $\Omega^{w=2}_{\pm}$ are the two $w=2$ twisted sector ground states.

In this paper we shall mainly concentrate on calculating the anomalous dimensions for the spin $h=1$ currents $W^3=\sqrt2\mathcal{M}^3$, as well as 
\begin{equation}
W^1 = \frac{\mathcal{M}^+ + \mathcal{M}^-}{\sqrt{2}} \ , \qquad
W^2 = \frac{\mathcal{M}^+ - \mathcal{M}^-}{\sqrt{2}i} \ ,
\end{equation}
which satisfy eq.~(\ref{Wnormalisation}). We shall also sometimes make sanity checks using the $\mathfrak{su}(2)$ R-symmetry currents: they should not acquire an anomalous dimension. As an illustration of our approach,
let us verify that this is the case for the Cartan generator ${\cal J}^3$ of the R-symmetry. In this case, we find 
\begin{equation}
\begin{aligned}
\bigl\langle \tilde\Phi(x_1) \Phi(x_2) &  \mathcal{J}^3(x)\mathcal{J}^3(y) \bigr\rangle =  
\braket{\tilde\Phi \Phi \mathcal{J}^3\mathcal{J}^3}\\
&=\sum_{n=0}^{\infty}
\left( \frac{\braket{\tilde\Phi \Phi(\mathcal{J}^3_n\mathcal{J}^3)}}{(x - y)^{n+1}}
+\frac{\braket{(\mathcal{J}^3_n\tilde\Phi)\Phi \mathcal{J}^3}}{(x-x_1)^{n+1}}
+\frac{\braket{\tilde\Phi(\mathcal{J}^3_n\Phi)\mathcal{J}^3}}{(x-x_2)^{n+1}} \right)+G(x)\ ,
\end{aligned}
\end{equation}
where the terms inside the brackets account for all the poles in $x$, while $G(x)$ is an entire function. (Here, the notation $(\mathcal{J}^3_nO)$ denotes the vertex operator corresponding to the state $\mathcal{J}^3_n |O\rangle$.) Since the $4$-point function goes as $x^{-2}$ as $x\rightarrow\infty$ while the terms in the sum go, at most, as $x^{-1}$, it follows that $G(x)= 0$ by Liouville's theorem. Since $\Phi$ and $\tilde{\Phi}$ are primary with respect to the R-symmetry currents, the only non-trivial contribution comes from the first term, and we obtain 
\begin{equation}\label{comp1}
\begin{aligned}
\braket{\tilde\Phi(x_1) \Phi(x_2) \mathcal{J}^3(x)\mathcal{J}^3(y)}=\frac{\braket{\tilde\Phi \Phi}}{2(x-y)^2}=\braket{\tilde\Phi(x_1) \Phi(x_2)}\braket{\mathcal{J}^3(x)\mathcal{J}^3(y)}=\frac{1}{2 (x_1-x_2)^2 (x - y)^2}\ .
\end{aligned}
\end{equation}
Since this is a disconnected correlator, it does not contribute to the logarithmic correction; in particular, therefore, it follows that $\mathcal{J}^3$ does not have an anomalous conformal dimension.

On the other hand, for the case where we replace $\Phi$ and $\tilde{\Phi}$ by $\Omega^{w=2}_{\pm}$, the same method yields
\begin{equation}
\begin{aligned}
\braket{\Omega^{w=2}_{+}(x_1)\Omega^{w=2}_{-}(x_2)\mathcal{J}^3(x)\mathcal{J}^3(y)}=\frac{1}{2 (x_1-x_2) (x - y)^2}+ \frac{1}{N} \, \frac{x_{1}-x_{2}}{4 (x_1 - x) (x_1 - y) (x_2 - x) (x_2 - y)} \ ,
\label{2.24}
\end{aligned}
\end{equation}
where the first term is again the disconnected contribution, while the second term comes from the connected correlator (and hence goes as $\frac{1}{N}$).\footnote{Technically, this factor arises because of the normalisation by $\frac{1}{\sqrt{N}}$ in eq.~(\ref{eq:map_to_untwisted_op}).} Later on, when we want to relate the worldsheet correlators to the spacetime correlators, we will also need the following spacetime correlator,
\begin{equation}
\begin{aligned}
\braket{\Omega^{w=2}_{+}(x_1)\Omega^{w=2}_{-}(x_2)\mathcal{J}^+(x)\mathcal{J}^-(y)}=\frac{1}{(x_1-x_2) (x - y)^2}+ \frac{1}{N} \, \frac{1}{(x_1 - y) (x_2 - x) (x - y)} \ .
\label{2.25}
\end{aligned}
\end{equation}
Let us remind the reader that with respect to the right-moving (antiholomorphic) part all of these correlators behave as 
\begin{equation}
\begin{aligned}
\braket{\bar{\tilde{\Phi}}(\bar x_1) \bar \Phi(\bar x_2)\Omega^1(\bar x)\Omega^1(\bar y)}=\braket{\bar{\tilde{\Phi}}(\bar x_1) \bar \Phi(\bar x_2)}=\frac{1}{(\bar x_1-\bar x_2)^2}\ ,
\label{2.22}
\end{aligned}
\end{equation}
and thus for the full correlator eqs.~(\ref{comp1}), (\ref{2.24}), and (\ref{2.25}) need to be multiplied by (\ref{2.22}).

\section{The worldsheet set-up} \label{sec:hyb-states}

In this section, we start our worldsheet computation by first explaining our notations for the hybrid string formalism. Our set-up is largely based on \cite{Dei:2020zui}, where more details can be found --- we mostly follow the same conventions. We then identify the physical states corresponding to the dual operators of interest, and formulate how to calculate their $4$-point correlators in the hybrid formalism.

\subsection{Generalities and the free field realisation}

In the hybrid formalism of Berkovits, Vafa and Witten (BVW) \cite{Berkovits:1999im}, string theory on the background $\text{AdS}_3 \times \text{S}^3 \times \mathbb{T}^4$ with pure NS-NS flux is described by three a priori uncoupled theories: \begin{enumerate*}[label=(\roman*)]
\item a $\mathfrak{psu}(1,1|2)_k$ WZW model at level $k$,
\item two (ghost) bosons $\rho$ and $\sigma$, and 
\item a topologically twisted $\mathbb{T}^4$ CFT (whose details are described in Appendix~\ref{app:N=4ofT4}).
\end{enumerate*}
The total theory has a twisted (small) $\mathcal{N}=(4,4)$ superconformal symmetry, and the physical  states $\bm{\Phi}$ are characterised by a double cohomology:
\begin{equation} \label{eq:physical}
G^+_0 \bm{\Phi} = \tilde{G}^+_0 \bm{\Phi} = (J_0-\tfrac{1}{2}) \bm{\Phi} = T_0 \bm{\Phi} = 0 \ , \qquad\qquad \bm{\Phi} \sim \bm{\Phi} + G^+_0 \tilde{G}^+_0 \bm{\Psi} \ ,
\end{equation}
where $G^+$, $\tilde{G}^+$, $J$ and $T$ are some of the $\mathcal{N}=4$ generators. (Here we have considered the left-moving states; a similar condition applies to the right-moving states.) At level $k=1$, the $\mathfrak{psu}(1,1|2)_1$ model admits a free field realisation in terms of four symplectic bosons ($\xi^{\pm}$ and $\eta^{\pm}$) and four free fermions ($\chi^{\pm}$ and $\psi^{\pm}$),
\begin{equation} \label{eq:opes-free fields}
\xi^{\alpha}(z) \eta^{\beta}(\zeta) \sim \epsilon^{\alpha\beta} \frac{1}{z-\zeta} \ , \qquad \qquad
\psi^{\alpha}(z) \chi^{\beta}(\zeta) \sim \epsilon^{\alpha\beta} \frac{1}{z-\zeta} \ ,
\end{equation}
where $\alpha, \beta \in \{\pm\}$, $\epsilon^{+-} = 1 = -\epsilon^{-+}$ and the other OPEs are trivial; the details of this are spelled out in Appendix~\ref{app:psu1}. For the time being it is worth stressing that the free fields actually only realise $\mathfrak{u}(1,1|2)_1$ rather than $\mathfrak{psu}(1,1|2)_1$. As a consequence, one cannot write the ${\cal N}=4$ generators of \cite{Berkovits:1999im} directly in terms of these free fields. However, this problem can be resolved by introducing additional ghosts \cite{Gaberdiel:2022als}, and this is also reviewed in Appendix~\ref{app:hybrid}.

\subsection{Marginal operators and spin 1 currents from the worldsheet} \label{sec:states}

In this subsection, we identify the worldsheet operators that are of interest to us. This is to say, we find the physical states of the worldsheet theory (i.e.\ the states that are characterised by eq.~(\ref{eq:physical})) that correspond to the various CFT fields we considered in Section~\ref{2}. 

Let us begin with the R-symmetry generators $\mathcal{J}^{a}$, whose worldsheet states were already identified in \cite{Gaberdiel:2021njm}
\begin{equation} \label{eq:ka-states}
\textbf{J}^a = K^a V^{w=1}_{\frac{1}{2},0} e^{2\rho+i\sigma+i H_{\mathbb{T}^4}} \ ,
\end{equation}
where $a\in\{3,\pm\}$. Here $V^{w=1}_{\frac{1}{2},0}$ denotes the state in the $w=1$ spectrally flowed sector whose ground state before spectral flow has $m_1=\frac{1}{2}$ and $m_2=0$, see  eq.~(\ref{eq:tw_ground_states}), and the $K^a$'s refer to the generators of $\mathfrak{su}(2)_1 \subset \mathfrak{psu}(1,1|2)_1$, see Appendix~\ref{app:psu1}.  
Throughout the text we use the convention that bold symbols describe worldsheet states, so as to distinguish them from those in the dual CFT. 	
\smallskip

Next we turn to the worldsheet states corresponding to the deformation operators.
In order to find them, we apply the DDF operators of \cite{Naderi:2022bus} to the worldsheet states that correspond to the $2$-cycle twisted sector ground states.\footnote{The DDF operators map physical states to physical states, and they satisfy the spacetime mode-algebra, so their consecutive application to an appropriate ground state yields the worldsheet physical state corresponding to any given state in the spacetime theory.} The $w$-cycle twisted ground states were already identified in Appendix~A.2 of \cite{Dei:2020zui}, however, only in the $P=-2$ picture. For reasons that will become clear later, we will also need them in other pictures. In fact, in the $\mathcal{N}=4$ description of string theory, there is a picture-raising operator that maps physical states to physical states, and that is defined as \cite{Blumenhagen:2013fgp}
\begin{equation}
P_+ = -G^+_0 (e^{-\rho-iH_{\mathbb{T}^4}})_0 \ ,
\end{equation}
see Appendix~\ref{app:hybrid}. Using this operator, we can find the $w$-cycle twisted sector ground states in the other pictures. For $w$ odd and in picture $P=-2n\leq 0$, it was found in \cite{Naderi:2022bus} that, up to some unimportant normalisation factor (here we have used the bosonised form for the fermions, see eq.~(\ref{eq:ferm-bosonisation}) in Appendix~\ref{app:C1}),
\begin{equation}
\bm{\Omega}^{w \text{ odd}}_{P=-2n}=(-w)^{1-n} e^{-(n-1)(iq_1-iq_2)} V^{w}_{m_1,m_2} e^{2n\rho+i\sigma+in H_{\mathbb{T}^4}} \ ,
\end{equation}
where
\begin{equation}
m_1 = -\frac{(w-1)^2}{8w}+\frac{n-1}{2} \ , \qquad\qquad m_2 = -\frac{(w+1)^2}{8w}-\frac{n-1}{2} \ .
\end{equation}
For $w$ even we find similarly (below, we put the $\pm$ index upstairs to avoid cluttering the symbol)
\begin{equation}
\bm{\Omega}^{w \text{ even}, \, \pm}_{P=-2n}=(\pm i w)^{1-n} e^{\pm \frac{iq_1+iq_2}{2}-(n-1)(iq_1-iq_2)} V^{w}_{m_1,m_2} e^{2n\rho+i\sigma+in H} \ ,
\end{equation}
where
\begin{equation}
m_1 = -\frac{(w-2)}{8}+\frac{n-1}{2} \ , \qquad\qquad m_2 = -\frac{(w+2)}{8}-\frac{n-1}{2} \ ,
\end{equation}
and the $\pm$ refers to the two different ground states for $w$ even. (Here $P_+ \bm{\Omega}^{w \text{ even}, \, \pm}_{P=-2n} = \bm{\Omega}^{w \text{ even}, \, \pm}_{P=-2(n-1)}$, as expected.)
\smallskip

In order to find the marginal operators, we need to apply the DDF operators of the supercurrents \cite{Naderi:2022bus,Gaberdiel:2022als}
\begin{subequations}
\begin{equation}\label{eq:ddf-g+}
\bm{G}^{+}_{-\frac{1}{2}} = \oint dz \Big( \chi^+ \xi^+ e^{-\rho} \tilde{G}^-_{\mathbb{T}^4} \Big)(z) \ ,
\end{equation}
\begin{equation} \label{eq:ddf-g-}
\bm{G}^{-}_{-\frac{1}{2}}=\oint dz \Big(-\chi^- \xi^+ e^{-\rho} G^-_{\mathbb{T}^4} + \chi^- \xi^+ e^{-\rho-i\sigma}+\psi^- \eta^+ \Big)(z) \ ,
\end{equation}
\end{subequations}
to the $2$-cycle twisted ground states $\bm{\Omega}^{2\pm}_{P=-2n}$. 
To end up with states of definite ghost number, i.e.\ states that have only a single term, we choose to apply $\bm{G}^{+}_{-\frac{1}{2}}$ to $\bm{\Omega}^{2-}_{P=0}$, and $\bm{G}^{-}_{-\frac{1}{2}}$ to 
$\bm{\Omega}^{2+}_{P=-6}$.\footnote{If the modes of $\tilde{G}^{\pm}$ are used instead, one gets the same states as the ones in eqs.~\eqref{eq:deformations} up to BRST-exact terms.} Here $P=-6$ is chosen for simplicity, because the second and third terms in eq.~(\ref{eq:ddf-g-}) vanish for $n\geq 3$. The other picture number, $P=0$, is then fixed by the requirement that the total picture number in the $2$-point function of the two deformation operators is $P=(-6+1)+(0+1)=-4$, as required by eq.~\eqref{eq:picture-condition} below. Explicitly, this then leads to 
\begin{subequations} \label{eq:deformations}
\begin{align}
\bm{\Phi} &= \bm{G}^{+}_{-\frac{1}{2}} \bm{\Omega}^{2-}_{P=0} = 2 i e^{\frac{3}{2}(iq_1-iq_2)} V^{w=2}_{-\frac{1}{2},\frac{1}{2}} e^{-\rho+i\sigma} \tilde{G}^-_{\mathbb{T}^4} \ , \\
\tilde{\bm{\Phi}} &= \bm{G}^{-}_{-\frac{1}{2}} \bm{\Omega}^{2+}_{P=-6} = -\frac{i}{4} e^{-\frac{3}{2}(iq_1-iq_2)} V^{w=2}_{1,-1} e^{5\rho+i\sigma+2iH_{\mathbb{T}^4}} \tilde{G}^+_{\mathbb{T}^4} \ .
\end{align}
\end{subequations}
\smallskip

Finally, we need the worldsheet states corresponding to the spin-$1$ currents, $\mathcal{M}^{a}$ with $a\in\{3,\pm\}$. To do so, we apply the DDF operators corresponding to the free fermions, see \cite[ (3.16)]{Naderi:2022bus}, to the CFT vacuum state $\bm{\Omega}^1_{(P=0)}$, mimicking the CFT formulae of eq.~(\ref{eq:other-su2}). We take the ground state to have $P=0$ and each of the fermions to have $P=-1$ so that the  resulting state is in the $P=-2$ picture; as we shall see, this will simplify the calculation of the correlation functions. The results are
\begin{subequations} \label{eq:states-other-su2}
\begin{align}
\textbf{M}^3&= \frac{1}{2} \Big( \partial\big(-iq_1+iq_2+2\rho+2iH^1_{\mathbb{T}^4}\big) V^{w=1}_{\frac{1}{2},
0} + 2 \xi^-_{-\frac{3}{2}} V^{w=1}_{1,0} \Big) e^{2\rho+i\sigma+iH_{\mathbb{T}^4}} \ , \\
\textbf{M}^+&=- V^{w=1}_{\frac{1}{2},0} e^{2\rho+i\sigma+2iH^1_{\mathbb{T}^4}} \ , \\
\textbf{M}^-&=+ V^{w=1}_{\frac{1}{2},0} e^{2\rho+i\sigma+2iH^2_{\mathbb{T}^4}} \ .
\end{align}
\end{subequations}

\subsection{Correlators in the hybrid formalism} \label{sec:hybrid_correlators}

With these preparations at hand, we now want to calculate the correlation functions of Section~\ref{sec:conf_pert_theory} on the worldsheet.  In the hybrid formalism, the unintegrated $n$-point functions on the sphere are \cite{Berkovits:1999im,Dei:2020zui,Berkovits:1994vy,Gaberdiel:2021njm}
\begin{equation} \label{eq:cor}
\left\langle \Big[\prod_{k=1}^{n-2} B(z_{n+k})\Big][F_0 \bm{\Phi}_1](z_1) \bm{\Phi}_2(z_2)(\tilde{G}^-_{-1} \bm{\Phi}_3)(z_3) \cdots (\tilde{G}^-_{-1} \bm{\Phi}_n)(z_n)\right\rangle \ ,
\end{equation}
where $\tilde{G}^-_{-1}$ is defined in eq.~(\ref{eq:gtilde-}), and $B$ is the field that was denoted by $W$ in \cite{Dei:2020zui} --- it corresponds to the vacuum state with respect to $\mathfrak{psu}(1,1|2)_1$, but it is required in order to balance the $U_0$ charge that is affected by the actions of $\tilde{G}^-_{-1}$, see \cite[Section~2.2]{Dei:2020zui} for more details. Here 
\begin{equation}
F_0 = (e^{i\sigma})_0 (e^{-\rho-iH_{\mathbb{T}^4}})_0 \ ,
\end{equation}
accounts for the fact that, according to \cite[ (3.8)]{Dei:2020zui}, we need 
$(\tilde{G}^+)^{-1}\sim e^{-\rho-iH_{\mathbb{T}^4}}$ and $(G^-)^{-1}\sim e^{i\sigma}$.
We shall assume that the picture numbers $P_i$ of the $\Phi_i$ in eq.~\eqref{eq:cor} satisfy \cite{Blumenhagen:2013fgp}
\begin{equation} \label{eq:picture-condition}
\sum_{i=1}^n P_i = - 2 n \ .
\end{equation}
For the physical states $\Phi_i$ at hand, the picture number is related to the $Y_0$ charge via \cite{Gaberdiel:2021njm}
\begin{equation}\label{pic}
P_i = -2 - y_i \ ,
\end{equation}
where $y_i$ is the eigenvalue of $\Phi_i$ under $Y_0=U_0-V_0$, see Appendix~\ref{app:psu1}. Since the $B$ fields have $Y_0$ charge $=2$ and $Z_0$ charge $=0$, the condition in eq.~(\ref{eq:picture-condition}) is therefore equivalent to 
\begin{equation}
\sum_{i=1}^n y_i = 0 \ . 
\end{equation}
This therefore just imposes that the total $Y_0$ charge of the operators in the correlator must vanish, see \cite{Gaberdiel:2021njm}.

In order to obtain the correlators in the dual CFT, we have to integrate eq.~\eqref{eq:cor} over the worldsheet moduli. In fact, because of the ${\rm SL}(2,\mathbb{C})$ symmetry, one can first fix the worldsheet $z_i$ positions 
of any three operators $\Phi_i$, and then integrate over the remaining insertion points (moduli).\footnote{Obviously, we can also use the M\"obius symmetry of the dual spacetime CFT, to set three of the $x_i$ positions to any fixed values.} We do not know the precise measure with which this integral will have to be performed, and so this will only fix our result up to some undetermined constants. We shall determine them below by comparison with other correlators.

\section{Worldsheet 4-point functions}\label{fix}

In this section we shall first compute the schematic $4$-point functions
\begin{equation} \label{eq:spin-1-d}
\begin{aligned}
\left< B B (F_0\tilde{\bm{\Phi}}) \bm{\Phi} (\tilde{G}^-_{-1}\textbf{J}^3) (\tilde{G}^-_{-1}\textbf{J}^3) \right> \ , \quad
\left< B B (F_0\tilde{\bm{\Phi}}) \bm{\Phi} (\tilde{G}^-_{-1}\textbf{J}^+) (\tilde{G}^-_{-1}\textbf{J}^-) \right> \ , \quad\quad \\
\left< B B (F_0\tilde{\bm{\Phi}}) \bm{\Phi} (\tilde{G}^-_{-1}\textbf{M}^3) (\tilde{G}^-_{-1}\textbf{M}^3) \right> \ , \quad 
\left< B B (F_0\tilde{\bm{\Phi}}) \bm{\Phi} (\tilde{G}^-_{-1}\textbf{M}^+) (\tilde{G}^-_{-1}\textbf{M}^-) \right> \ ,
\end{aligned}
\end{equation}
including their overall normalisation. (This will be fixed by comparison of some related correlators to the dual CFT answer.) We shall then integrate these $4$-point functions over the worldsheet moduli and thus determine the corresponding dual CFT correlators.  In the following, we shall spell out the details for one case explicitly, namely for 
\begin{equation} \label{eq:fixing-cor}
\braket{B(z_5)\, B(z_6)\, (F_0\tilde{\bm{\Phi}})(z_1,x_1)\bm{\Phi}(z_2,x_2)(\tilde{G}^-_{-1}\textbf{J}^3)(z_3,x_3)(\tilde{G}^-_{-1}\textbf{J}^3)(z_4,x_4)} \ ,
\end{equation}
where we have reinstated the insertion points on the worldsheet ($z_i$), and the dual CFT ($x_i$) for clarity. The other cases can be dealt with similarly, and further details are given in Appendix~\ref{WW}. In order to fix the various undetermined functions, see below, we shall also calculate the corresponding correlators for the case where we replace the left-moving part of the perturbing fields by the ground states in the $2$-cycle twisted sector, i.e. 
\be
\tilde{\bm{\Phi}} \ \mapsto \ \bm{\Omega}^{w=2,+}_{P=-2} \ , \qquad \qquad
\bm{\Phi}\  \mapsto \ \bm{\Omega}^{w=2,-}_{P=-2}  \ . 
\ee

The above correlators only describe the left-moving part on the worldsheet. Since the currents we are considering are all left-moving in spacetime, i.e.\ involve the vacuum on the right, their right-moving component on the worldsheet is the state $\bm{\Omega}^{1}$, corresponding to the spacetime vacuum. Thus the right-moving part of the worldsheet correlators is in all cases the same, namely 
\begin{equation} \label{eq:spin-1-d-right}
\left< \overline{B} (\bar z_5) \overline{B} (\bar z_6) (\overline{F_0{\tilde{\bm{\Phi}}}})(\bar z_1, \bar x_1) \overline{\bm{\Phi}} (\bar z_2, \bar x_2) (\overline{\tilde G^-_{-1}\bm{\Omega}^{1}})(\bar z_3, \bar x_3) (\overline{\tilde G^-_{-1}\bm{\Omega}^{1}})(\bar z_4, \bar x_4) \right>  \ . 
\end{equation}

\subsection{The calculation of eq.~(\ref{eq:fixing-cor})} \label{sec:c}

To start with, let us note that the sum of the picture numbers is $\sum P_i = -8=-2\times 4$, as required by eq.~(\ref{eq:picture-condition}). Moreover, in $\tilde{G}^-_{-1}$ there are several terms with different ghost contributions, see eq.~(\ref{eq:gtilde-}). We observe that the correct background charge of the ghost $\rho$, which is $\Lambda_{\rho}=3$, is already accounted for by the fields $F_0 \tilde{\bm{\Phi}} = (e^{i\sigma})_0 (e^{-\rho-iH})_0 \tilde{\bm{\Phi}}$ and $\bm{\Phi}$, see eq.~(\ref{eq:t-hybrid}), and hence the remaining terms should lead to a trivial $\rho$ exponent. Thus, only one term in $\tilde{G}^-_{-1}\textbf{J}^3$ can contribute, and we end up with the correlator 
\begin{equation} \label{eq:cor-k3k3}
\left< B B  [F_0 \tilde{\bm{\Phi}}] \bm{\Phi} (e^{iq_1-iq_2} K^3 V^{w=1}_{0,\frac{1}{2}}) (e^{iq_1-iq_2} K^3 V^{w=1}_{0,\frac{1}{2}}) \right> \ ,
\end{equation}
where $(e^{iq_1-iq_2} K^3 V^{w=1}_{0,\frac{1}{2}})$ is the normal ordered product of these three fields; since $K^3$ commutes with $e^{iq_1-iq_2}$ this normal-ordering is unambiguous. 

Given the structure of the worldsheet theory, the correlator factorises into contributions coming from the symplectic bosons, the fermions, the $\rho$ and $\sigma$ bosons, and the topologically twisted $\mathbb{T}^4$, and we can calculate them independently from one another. 
Let us start with the symplectic boson contribution to the correlator, 
\begin{equation} \label{eq:k3k3-symp}
\left<B(z_5)\, B (z_6)\, V^{w=2}_{1,-1}(z_1,x_1) V^{w=2}_{-\frac{1}{2},\frac{1}{2}}(z_2,x_2) V^{w=1}_{0,\frac{1}{2}}(z_3,x_3) V^{w=1}_{0,\frac{1}{2}}(z_4,x_4) \right> \ .
\end{equation}
Using the incidence relation of \cite{Dei:2020zui} we can express this correlator as, see Appendix~\ref{WW} for more details, 
\begin{equation}\label{sympbc}
C(x_l,\bar{x}_l,z_l,\bar{z}_l)\, \delta^{(2)}(x_4 - \Gamma(z_4)) \, \prod_{i=1}^4a_i^{-2m_1^i}b_i^{2(m_1^i-m_2^i)} \ .
\end{equation}
In writing this formula we have combined the correlator with the corresponding right-moving term, see eq.~(\ref{eq:spin-1-d-right}), which involves the same spectrally flowed sectors and hence exhibits the same localisation behaviour; these terms combine to give a two-dimensional delta-function. The undetermined normalisation factor $C(x_l,\bar{x}_l,z_l,\bar{z}_l)$ depends on the spectral flow sectors (and the insertion points), but not on their $(m_1,m_2)$ quantum numbers.\footnote{As is explained in Appendix~\ref{WW}, the correlator is only non-zero provided that the $U_0$-charge condition ${\cal S}=0$, see eq.~(\ref{Sdef}), is satisfied.} It will therefore be the same for all the different correlators we will consider in this paper. 
\smallskip

The contribution from the fermions of $\mathfrak{u}(1,1|2)_1$ is (in bosonised form)
\begin{equation} \label{eq:u-ferm}
\begin{aligned} 
\Bigl\langle (e^{-(iq_1-iq_2)})(z_5) \, (e^{-(iq_1-iq_2)})(z_6) &  \, (e^{-\frac{3}{2}(iq_1-iq_2)})(z_1) \, (e^{\frac{3}{2}(iq_1-iq_2)})(z_2) \\
& \qquad \times (K^3 e^{iq_1-iq_2})(z_3) \, (K^3 e^{iq_1-iq_2})(z_4) \Bigr\rangle \ ,
\end{aligned}
\end{equation}
where $K^3$ equals $\frac{1}{2}\partial(iq_1+iq_2)$. If we write $\phi=q_1-q_2$ and $\kappa=q_1+q_2$, then the OPE of $\phi$ with $\kappa$ is trivial, while 
\be
\phi(z) \phi(\zeta) \sim - 2 \ln(z-\zeta) \ , \qquad\qquad \kappa(z) \kappa(\zeta) \sim - 2 \ln(z-\zeta)
\ee
and neither $\phi$ nor $\kappa$ has any background charge. The above correlator can then be calculated using eq.~(\ref{eq:cor-exp}), and the result is
\be
\displaystyle e^{\frac{i\pi}{4}} \frac{(z_5-z_6)^2 (z_5-z_1)^3 (z_6-z_1)^3 (z_2-z_3)^3 (z_2-z_4)^3}{2 (z_5-z_2)^3 (z_5-z_3)^2 (z_5-z_4)^2 (z_6-z_2)^3 (z_6-z_3)^2 (z_6-z_4)^2 (z_1-z_2)^{9/2} (z_1-z_3)^3 (z_1-z_4)^3} \ .
\ee
Note that one needs to be careful about the cocycle factors (that guarantee that the exponentials in eq.~(\ref{eq:ferm-bosonisation}) define fermions), but this at most affects the answer by a relative factor of modulus one. 
\smallskip

The contribution from the $\rho$ and $\sigma$ bosons equals 
\begin{equation} \label{eq:k3k3-rho-sigma}
\left< (e^{4\rho+2i\sigma})(z_1) (e^{-\rho+i\sigma})(z_2) \right>  = (z_1 - z_2)^6 \ ,
\end{equation}
where we have used eq.~(\ref{eq:cor-exp}) together with $\Lambda_{\rho}=\Lambda_{\sigma}=3$.
Finally, for the contribution from the (twisted) $\mathbb{T}^4$, see Appendix~\ref{app:N=4ofT4} for our conventions, we can calculate the bosonic and fermionic contributions separately, and this leads to 
\begin{equation}
\left< (e^{iH} \tilde{G}^+_{\mathbb{T}^4})(z_1) (\tilde{G}^-_{\mathbb{T}^4})(z_2) \right>  = \frac{2}{(z_1-z_2)^2} \ .
\end{equation}

\subsection{Fixing the overall normalisation factor}\label{sec:normalisation}

As already alluded to before, we only know how to determine the symplectic boson correlators up to the normalisation factor $C(x_l,\bar{x}_l,z_l,\bar{z}_l)$ of eq.~(\ref{sympbc}). However, since the same normalisation factor appears in all the correlators we will consider in this paper, we can fix it by comparing our worldsheet calculation with known boundary correlators, and this is the strategy we shall pursue in the following. 
In doing so, there are two subtleties that we need to be careful about. First of all, we cannot directly read off the  worldsheet correlator (and in particular the function  $C(x_l,\bar{x}_l,z_l,\bar{z}_l)$) from this perspective, since it is the {\em integrated} worldsheet correlator (where we integrate over the worldsheet variables $z_i$ and $\bar{z}_i$) that is to be identified with the dual CFT correlator. This modular integral is only well-defined as a two-dimensional worldsheet integral (and normalisations 
only make sense for the full correlator), and thus we need to combine the left- and right-moving components before doing the integral. 

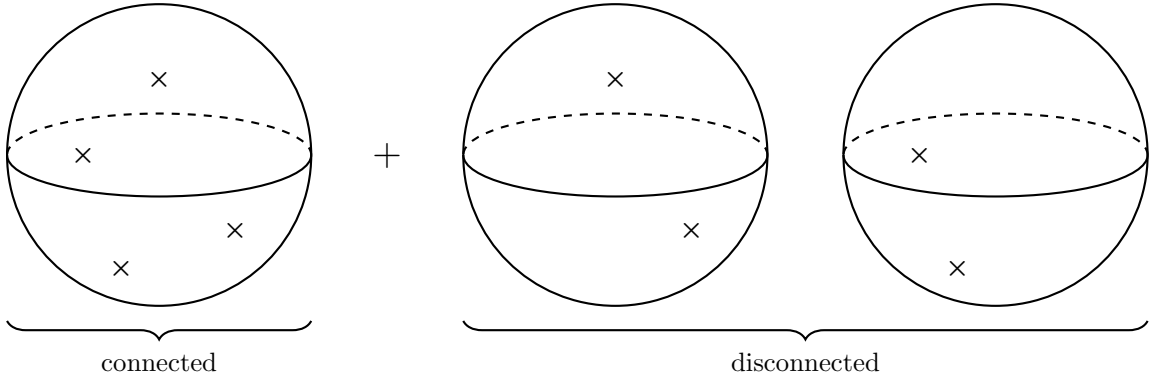
\begin{figure} 
\centering
\begin{tikzpicture}
\begin{scope}
\draw[thick] (0,0) circle (2);
\draw[thick] (0,0) [partial ellipse = 0:-180:2 and 0.55];
\draw[thick, dashed] (0,0) [partial ellipse = 0:180:2 and 0.55];
\node at (0,1) {$\boldsymbol{\times}$};
\node at (1,-1) {$\boldsymbol{\times}$};
\node at (-1,0) {$\boldsymbol{\times}$};
\node at (-0.5,-1.5) {$\boldsymbol{\times}$};
\end{scope}
\node at (3,0) {\Large $+$};
\begin{scope}[xshift = 6cm]
\draw[thick] (0,0) circle (2);
\draw[thick] (0,0) [partial ellipse = 0:-180:2 and 0.55];
\draw[thick, dashed] (0,0) [partial ellipse = 0:180:2 and 0.55];
\node at (0,1) {$\boldsymbol{\times}$};
\node at (1,-1) {$\boldsymbol{\times}$};
\end{scope}
\begin{scope}[xshift = 11cm]
\draw[thick] (0,0) circle (2);
\draw[thick] (0,0) [partial ellipse = 0:-180:2 and 0.55];
\draw[thick, dashed] (0,0) [partial ellipse = 0:180:2 and 0.55];
\node at (-1,0) {$\boldsymbol{\times}$};
\node at (-0.5,-1.5) {$\boldsymbol{\times}$};
\end{scope}
\draw[thick, decorate, decoration = {brace, amplitude=7pt}] (2,-2.2) -- (-2,-2.2);
\node[below] at (0,-2.5) {connected};
\draw[thick, decorate, decoration = {brace, amplitude=7pt}] (13,-2.2) -- (4,-2.2);
\node[below] at (8.5,-2.5) {disconnected};
\end{tikzpicture}
\caption{Connected and disconnected worldsheet contributions to the spacetime $4$-point functions considered in this paper.}
\label{fig}
\end{figure}

Secondly, there are in general different worldsheet topologies that contribute to a given dual CFT correlator --- in particular, for disconnected dual CFT correlators we also need to consider the contribution where we have two spherical worldsheets  --- and we need to be careful about disentangling the different contributions. In our case where we consider a $4$-point function and include only the first $g_s$ correction,  there are two possible configurations: either all four points lie on the same worldsheet sphere (connected), or we have two worldsheet spheres, each containing two points (disconnected), see Figure~\ref{fig}.\footnote{We thank Bob Knighton for providing  us with this figure.} In terms of the dual CFT, the disconnected term will contribute at order ${\cal O}(1)$, while the connected term contributes at order ${\cal O}(\frac{1}{N})$. 

The three dual CFT correlators that we shall use to determine the normalisation factors 
are the ones given in eqs.~(\ref{comp1}), (\ref{2.24}), (\ref{2.25}), all multiplied by \eqref{2.22}. All of them have a term that is of order ${\cal O}(1)$ in the large $N$ limit, while eqs.~(\ref{2.24}) and (\ref{2.25}) also have a contribution at order ${\cal O}(\frac{1}{N})$. The ${\cal O}(1)$ contribution reflects that all the three correlators contain a disconnected part; this will be reproduced from the worldsheet perspective by the contribution involving two worldsheet spheres, each containing only two points. Thus, schematically, we expect to find up to order $\frac{1}{N}$, 
\begin{align}
\braket{\tilde\Phi \Phi \mathcal{J}^3\mathcal{J}^3} & = \frac{1}{2 |x_1-x_2|^4 (x - y)^2} \nonumber \\
& = A \, \frac{1}{2 |x_1-x_2|^4(x - y)^2} + g_s^2 \int_{\mathbb{C}} d^2 z_4  \braket{BB (F_0\tilde{\bm{\Phi}})\bm{\Phi} (\tilde{G}^-_{-1} \textbf{J}^3) (\tilde{G}^-_{-1} \textbf{J}^3)}_{\text{local}} \ , \label{test1}
\end{align}
\begin{align}
& \braket{\Omega^{w=2}_{+}\Omega^{w=2}_{-}\mathcal{J}^3\mathcal{J}^3}  \nonumber \\
& \qquad = \frac{1}{2 (x_1-x_2) (\bar x_1-\bar x_2)^2(x - y)^2}
+ \frac{1}{N} \, \frac{x_{1}-x_{2}}{4 (x_1 - x) (x_1 - y) (x_2 - x) (x_2 - y)(\bar x_1-\bar x_2)^2}  \label{test2} \\ 
& \qquad = A \, \frac{1}{2 (x_1-x_2) (\bar x_1-\bar x_2)^2(x - y)^2} + g_s^2  \int_{\mathbb{C}} d^2 z_4   \braket{BB (F_0\bm{\Omega}^{w=2}_{+}) \bm{\Omega}^{w=2}_{-}(\tilde{G}^-_{-1} \textbf{J}^3) (\tilde{G}^-_{-1} \textbf{J}^3)}_{\text{local}} \  ,\nonumber 
\end{align}
and 
\begin{align}
& \braket{\Omega^{w=2}_{+}\Omega^{w=2}_{-}\mathcal{J}^+\mathcal{J}^-} \nonumber \\
& \qquad = \frac{1}{(x_1-x_2)(\bar x_1-\bar x_2)^2 (x - y)^2}+ \frac{1}{N} \, \frac{1}{(x_1 - y) (x_2 - x) (x - y)(\bar x_1-\bar x_2)^2}  \label{test3} \\ 
& \qquad = A \, \frac{1}{ (x_1-x_2)(\bar x_1-\bar x_2)^2 (x - y)^2} + g_s^2  \int_{\mathbb{C}} d^2 z_4   \braket{BB (F_0\bm{\Omega}^{w=2}_{+}) \bm{\Omega}^{w=2}_{-}(\tilde{G}^-_{-1} \textbf{J}^+) (\tilde{G}^-_{-1} \textbf{J}^-)}_{\text{local}} \ , \nonumber
\end{align}
where the subscript \emph{local} is meant to emphasise that these are the local correlators, containing both left- and right-moving components. We have also introduced a normalisation factor $A$ for the leading term involving two worldsheet spheres --- this should go as $A = 1 + {\cal O}(\frac{1}{N})$ --- while $g_s^2 \sim \frac{1}{N}$.\footnote{For the disconnected term each worldsheet describes a $2$-point function, and thus the correlator is completely fixed, up to normalisation by the conformal symmetry.} We now want to find exact expressions for $A$, $g_s^2$ and the unknown normalisation function $C(x_l,\bar{x}_l,z_l,\bar{z}_l)$ so that eqs.~(\ref{test1}), (\ref{test2}) and (\ref{test3}) hold to order $g_s^2 \sim \frac{1}{N}$. 
In order to do so, we need to calculate the above integrals. For the one appearing in eq.~(\ref{test1}) the integrand takes the form 
\begin{equation}
\begin{aligned}
&\braket{BB (F_0\tilde{\bm{\Phi}})\bm{\Phi} (\tilde{G}^-_{-1} \textbf{J}^3) (\tilde{G}^-_{-1} \textbf{J}^3)}_{\text{local}}=  \tilde C \, \delta^{(2)}(\Gamma(z_4)-x_4) \ ,
\label{ddkkws}
\end{aligned}
\end{equation}
where $\tilde C$ equals 
\begin{equation}
\begin{aligned}\label{Ctilde}
 \frac{1}{2 z_3^3}\left( \frac{e^{\frac{3\pi}{4} i}\ (z_5-z_6)^2}{(z_5-1)(z_6-1)z_5z_6(x_1-x_3)(x_3-x_2)} \right)\left( \frac{(x_2-x_1)z_4^2+(x_3-x_2)(2z_4-1)}{z_4^{1/2}(z_4-1)^{1/2} \, (z_4-z_5)(z_4-z_6)} \right) \!
 C'(x_l,\bar{x}_l,z_l,\bar{z}_l)  ,
\end{aligned}
\end{equation}
and $C'(x_l,\bar{x}_l,z_l,\bar{z}_l)$ differs from $C(x_l,\bar{x}_l,z_l,\bar{z}_l)$, defined in eq.~\eqref{sympbc}, by the contribution from the right-moving fermions, ghosts and $\mathbb{T}^4$ excitations, see eq.~\eqref{eq:spin-1-d-right}.
Here we have used the ${\rm SL}(2,\mathbb{C})$ invariance in eq.~\eqref{ddkkws} to set $z_1=0$, $z_2=1$, while $z_3\rightarrow \infty$, and we have only written out the terms that survive in this limit. With this convention the relevant covering map is 
\begin{equation}
\begin{aligned}
\Gamma(z)=\frac{x_3(x_2-x_1)z^2-x_1(x_2-x_3)(2z-1)}{(x_2-x_1)z^2-(x_2-x_3)(2z-1)}\ .
\end{aligned}
\end{equation}
\smallskip

\noindent Similarly, the correlator that appears in eq.~(\ref{test2}), is 
\begin{equation}
\begin{aligned}
&\braket{BB (F_0\bm{\Omega}^{w=2}_{+}) \bm{\Omega}^{w=2}_{-}(\tilde{G}^-_{-1} \textbf{J}^3) (\tilde{G}^-_{-1} \textbf{J}^3)}_{\text{local}}=\frac{(1-2z_4+2z_4^2)}{2z_4(z_4-1)}(x_1-x_2)\, \tilde C\, \delta^{(2)}(\Gamma(z_4)-x_4)\ ,
\label{vvkkws}
\end{aligned}
\end{equation}
while for the integrand of eq.~(\ref{test3}) we find 
\begin{equation}
\begin{aligned}
&\braket{BB (F_0\bm{\Omega}^{w=2}_{+}) \bm{\Omega}^{w=2}_{-}(\tilde{G}^-_{-1} \textbf{J}^+) (\tilde{G}^-_{-1} \textbf{J}^-)}_{\text{local}}= \frac{2(z_4-1)}{z_4}(x_1-x_2)\, \tilde C\, \delta^{(2)}(\Gamma(z_4)-x_4)\ .
\label{vvkkws1}
\end{aligned}
\end{equation}

These worldsheet correlators now need to be integrated over the remaining modulus, which is $z_4$ (and $\bar{z}_4$)  in our case. There are two solutions to the equation $\Gamma(z_4)=x_4$, namely 
\begin{equation}
\begin{aligned}
&z^{\pm}_4=\frac{\sqrt{(x_1-x_4)(x_2-x_3)}(\sqrt{(x_1-x_4)(x_2-x_3)}\pm\sqrt{(x_1-x_3)(x_2-x_4)})}{(x_2-x_1)(x_3-x_4)} \ .
\label{soln}
\end{aligned}
\end{equation}
For example for eq.~(\ref{ddkkws}) this leads to
\begin{align}
\int d^2 z_4  \braket{BB (F_0\tilde{\bm{\Phi}})\bm{\Phi} (\tilde{G}^-_{-1} \textbf{J}^3) (\tilde{G}^-_{-1} \textbf{J}^3)}_{\text{local}} & = 
\left.\frac{\tilde C}{|\Gamma'(z_4)|^2}\right|_{z_4^+}+\left.\frac{\tilde C}{|\Gamma'(z_4)|^2}\right|_{z_4^-} \ , \label{4.22} 
\end{align}
and similarly in the other two cases.

Plugging in the different terms, we find that the dual CFT answers, namely eqs.~(\ref{comp1}), (\ref{2.24}), and (\ref{2.25}), are matched provided that
\be\label{iden1}
A=1-\frac{2}{N} \ , 
\ee
and the terms $C^{\pm}\equiv\left.\frac{\tilde C}{|\Gamma'(z_4)|^2}\right|_{z_4^{\pm}}$ that depend on $C(x_l,\bar{x}_l,z_l,\bar{z}_l)$ take the form 
\begin{align}
g_s^2C^{\pm}= &\pm\frac{\left( 2(x_1-x_4)(x_2-x_3)+(x_1-x_2)(x_3-x_4)\pm2\sqrt{(x_1-x_4)(x_2-x_3)(x_1-x_3)(x_2-x_4)} \right)}{4N\sqrt{(x_1-x_3)(x_1-x_4)(x_2-x_3)(x_2-x_4)}|x_1-x_2|^4(x_3-x_4)^2}  \ .
\label{cpmfinal}
\end{align}

\subsection{Evaluating the dual CFT correlators}\label{sec:4.3}

Having fixed the normalisations we are now ready to integrate the full $4$-point functions over the worldsheet moduli and thus determine the dual CFT correlators involving the spin $1$ fields ${\cal M}^a$ with $a\in\{3,\pm\}$. More specifically, we need to calculate 
\begin{equation}\label{eq:spin-1-d-5}
\left< B B (F_0\tilde{\bm{\Phi}}) \bm{\Phi} (\tilde{G}^-_{-1}\textbf{M}^+) (\tilde{G}^-_{-1}\textbf{M}^-) \right>_{\text{local}} \ ,
\qquad
\left< B B (F_0\tilde{\bm{\Phi}}) \bm{\Phi} (\tilde{G}^-_{-1}\textbf{M}^3) (\tilde{G}^-_{-1}\textbf{M}^3) \right>_{\text{local}} \ ,
\end{equation}
where the $\textbf{M}^{a}$ are defined in eq.~(\ref{eq:states-other-su2}).  Let us begin with the first one.
The unintegrated correlation function equals
\begin{equation}\label{5.4}
\tilde{C}\,  \frac{1-2 z_4+2z_4^2}{z_4(z_4-1)}\, \delta^{(2)}(\Gamma(z_4)-x_4) \ ,
\end{equation}
where $\tilde{C}$ is given in eq.~(\ref{Ctilde}). Performing the integral and using eq.~(\ref{cpmfinal}) we then obtain
\begin{equation}\label{5.5}
\begin{split}
g_s^2\, & \int_{\mathbb{C}} d^2 z_4 \left< B B  (F_0\tilde{\bm{\Phi}}) \bm{\Phi} (\tilde{G}^-_{-1}\textbf{M}^+) (\tilde{G}^-_{-1}\textbf{M}^-) \right>_{\text{local}} \\
& \qquad = \frac{2}{N} \Bigl[ \frac{1}{|x_1-x_2|^4(x_3-x_4)^2}+\frac{1}{4(x_1-x_3)(x_1-x_4)(x_2-x_3)(x_2-x_4)(\bar x_1-\bar x_2)^2} \Bigr] .
\end{split}
\end{equation}

The calculation of the second $4$-point function in eq.~(\ref{eq:spin-1-d-5}) requires more effort, since the symplectic boson part involves the terms 
\begin{equation}
\begin{aligned}
\left< \tilde{B}(z_5) \tilde{B}(z_6) V^{w=2}_{1,-1}(z_1,x_1) V^{w=2}_{-\frac{1}{2},\frac{1}{2}}(z_2,x_2) [\xi^-_{-\frac{3}{2}} V^{w=1}_{\frac{1}{2},\frac{1}{2}}](z_3,x_3) V^{w=1}_{0,\frac{1}{2}}(z_4,x_4) \right> \ , \\
\left< \tilde{B}(z_5) \tilde{B}(z_6) V^{w=2}_{1,-1}(z_1,x_1) V^{w=2}_{-\frac{1}{2},\frac{1}{2}}(z_2,x_2) V^{w=1}_{0,\frac{1}{2}}(z_3,x_3) [\xi^-_{-\frac{3}{2}} V^{w=1}_{\frac{1}{2},\frac{1}{2}}](z_4,x_4) \right> \ , \\
\left< \tilde{B}(z_5) \tilde{B}(z_6) V^{w=2}_{1,-1}(z_1,x_1) V^{w=2}_{-\frac{1}{2},\frac{1}{2}}(z_2,x_2) [\xi^-_{-\frac{3}{2}} V^{w=1}_{\frac{1}{2},\frac{1}{2}}](z_3,x_3) [\xi^-_{-\frac{3}{2}} V^{w=1}_{\frac{1}{2},\frac{1}{2}}](z_4,x_4) \right> \ ,
\end{aligned}
\end{equation}
where the definition of $\tilde B$ can be found in eq.~\eqref{A11}. In order to evaluate these correlators we use the usual contour deformation techniques to turn the negative $\xi^-$ modes into non-negative modes, which are easier to deal with using eq.~\eqref{eq:Rgs} in particular. This leads to the (unintegrated) correlator 
\begin{equation}
\tilde{C} \frac{1-2 z_4+2z_4^2}{2 z_4(z_4-1)} \, \delta^{(2)}(\Gamma(z_4)-x_4) \ .
\end{equation}
Notice that this correlator is exactly half of eq.~\eqref{5.4}. We therefore immediately conclude that
\begin{equation}
\begin{split}
g_s^2 & \int_{\mathbb{C}} d^2 z_4 \left< B B (F_0\tilde{\bm{\Phi}}) \bm{\Phi} (\tilde{G}^-_{-1}\textbf{M}^3) (\tilde{G}^-_{-1}\textbf{M}^3) \right>_{\text{local}} \\
& \qquad = \frac{2}{N} \Bigl[ \frac{1}{2|x_1-x_2|^4(x_3-x_4)^2}+\frac{1}{8(x_1-x_3)(x_1-x_4)(x_2-x_3)(x_2-x_4)(\bar x_1-\bar x_2)^2}  \Bigr] \ .
\end{split}
\end{equation}
We have also checked by similar methods that
\begin{equation}\label{5.7}
\begin{split}
g_s^2\left< B B (F_0\tilde{\bm{\Phi}}) \bm{\Phi} (\tilde{G}^-_{-1}\textbf{J}^+) (\tilde{G}^-_{-1}\textbf{J}^-) \right>_{\text{local}} & = 
2 \cdot g_s^2\left< B B (F_0\tilde{\bm{\Phi}}) \bm{\Phi} (\tilde{G}^-_{-1}\textbf{J}^3) (\tilde{G}^-_{-1}\textbf{J}^3) \right>_{\text{local}} \ .
\end{split}
\end{equation}
Thus the correlator of eq.~(\ref{5.7}) only has a disconnected term after the $(z_4,\bar{z}_4)$ integration, and together with the contributions from the two worldsheets, c.f.\ eq.~(\ref{test2}) with $A=1-\frac{2}{N}$, this reproduces twice of eq.~(\ref{comp1}). In particular,  the corresponding currents $\mathcal{J}^a$ of the dual CFT therefore do not acquire an anomalous dimension, as expected. 

\section{The perturbation analysis for the spin 1 fields}\label{5}

With these results at hand, we can now determine the anomalous spacetime dimension of the ${\cal M}^a$ fields, following essentially the computation of \cite{Gaberdiel:2015uca}.

\subsection{Anomalous dimensions from 4-point functions}\label{sec:5.1}

 As was explained in Section~\ref{sec:conf_pert_theory}, only the connected part of the CFT amplitude contributes to the anomalous dimension. This can now be read off from eq.~(\ref{5.5})
\begin{subequations}
\begin{align}
\braket{\tilde{\bm{\Phi}}(x_1)\bm{\Phi}(x_2)\textbf{M}^+(x_3)\textbf{M}^-(x_4)}_{\rm conn} & = 
\frac{1}{2N(x_1-x_3)(x_1-x_4)(x_2-x_3)(x_2-x_4)(\bar x_1-\bar x_2)^2} \ , \\ 
\braket{\tilde{\bm{\Phi}}(x_1)\bm{\Phi}(x_2)\textbf{M}^3(x_3)\textbf{M}^3(x_4)}_{\rm conn}& 
=\frac{1}{4N(x_1-x_3)(x_1-x_4)(x_2-x_3)(x_2-x_4)(\bar x_1-\bar x_2)^2} \ ,
\end{align}
\end{subequations}
where we have used eq.~(\ref{iden1}) and eq.~(\ref{cpmfinal}). 

To obtain the anomalous dimension from this, one needs to integrate with respect to $x_1, x_2$ and then extract the coefficient of the log term $\ln|x_3-x_4|$.
In all cases the relevant integral is 
\begin{equation}
\begin{aligned}
\int d^2x_1d^2x_2 \frac{1}{(x_1-x_3)(x_1-x_4)(x_2-x_3)(x_2-x_4)}\frac{1}{(\bar x_1-\bar x_2)^2} \ ,
\label{int1}
\end{aligned}
\end{equation}
where we have taken the measure to be $d^2x_i= d\text{Re}(x_i)\, d\text{Im}(x_i)$. Using that 
$\frac{1}{(\bar x_1-\bar x_2)^2}=\bar\partial_2\frac{1}{(\bar x_1-\bar x_2)}$, as well as Stokes' theorem 
\begin{equation}
\begin{aligned}
\int_{A} d^2z\, \bar\partial f(z,\bar z)=\frac{1}{2i}\oint_{\partial A} dz f(z,\bar z) \ , 
\label{id}
\end{aligned}
\end{equation}
eq.~\eqref{int1} becomes
\begin{equation}
\begin{aligned}
&\int d^2x_1\frac{1}{2i(x_1-x_3)(x_1-x_4)}\oint dx_2 \frac{1}{(x_2-x_3)(x_2-x_4)(\bar x_1-\bar x_2)}\\
&
\qquad =\int d^2x_1\frac{1}{2i(x_1-x_3)(x_1-x_4)}(-2\pi i)\left( \frac{1}{(x_3-x_4)(\bar x_1-\bar x_3)}-\frac{1}{(x_3-x_4)(\bar x_1-\bar x_4)} \right) \ . \\
\label{int2}
\end{aligned}
\end{equation}
Note that the minus sign on the right-hand-side arises because the integration contour of $x_2$ around $x_2=x_3$ and $x_2= x_4$ is clockwise, as follows from eq.~\eqref{id}.
Next, since $\bar\partial_1\ln|(x_1-x_i)|^2=\tfrac{1}{(\bar x_1-\bar x_i)}$, we obtain\footnote{We are ignoring here the divergent contribution that comes from the singularity at $x_1=x_3$ and $x_1=x_4$ in the first and second term, respectively. This divergence can be thought of as a UV divergence and it needs to be regulated. However, it does not affect our conclusion because it leads to a term of the form $(x_{3} - x_4)^{-2}$ which does not contribute to the anomalous dimension.}

\begin{equation}
\begin{aligned}
\int d^2x_1\frac{-\pi}{(x_1-x_3)(x_1-x_4)}\left( \frac{\bar\partial_1\ln|(x_1-x_3)|^2}{(x_3-x_4)}-\frac{\bar\partial_1\ln|(x_1-x_4)|^2}{(x_3-x_4)} \right) &=\frac{-4\pi^2\ln|x_3-x_4|}{(x_3-x_4)^2} \ .
\label{int3}
\end{aligned}
\end{equation}
Applying this to the two cases we therefore conclude that the coefficients of $\ln|x_3-x_4|/(x_3-x_4)^2$, equal
\begin{equation}
\textbf{M}^+\textbf{M}^- : ~ - \frac{ 2\pi^2}{N} \ ,
\qquad\qquad
\textbf{M}^3\textbf{M}^3 : ~ - \frac{\pi^2}{N} \ .
\end{equation}
The relative factor of $2$ just reflects our normalisation convention for the currents, i.e.\ that 
\be
\langle(F_0\textbf{M}^3)\textbf{M}^3\rangle=\tfrac{1}{2}\langle (F_0\textbf{M}^+)\textbf{M}^-\rangle \ .
\ee
Thus all three currents have the same anomalous dimension,
 \begin{equation}
\lambda=\frac{g^2_\text{CFT}\pi^2}{4N}\ ,
\label{anom}
\end{equation}
where the factor of $g^2_\text{CFT}/2$ arises from the perturbation expansion --- $g_\text{CFT}$ denotes the coupling constant, see eq.~\eqref{2.15} --- and we have divided by $-4$, as discussed at the end of Section~\ref{sec:conf_pert_theory}.
This anomalous dimension then reproduces correctly the dual CFT answer. (The answer differs by a factor of $2$ relative to \cite{Gaberdiel:2015uca}, as was already pointed out in \cite{Apolo:2022fya}. This is because in \cite{Gaberdiel:2015uca}, the coefficient of $\ln(\bar x_3-\bar x_4)$ was considered, but one needs to match the coefficients of $\ln|x_3-x_4|$ and not separately $\ln(\bar x_3-\bar x_4)$ or $\ln (x_3-x_4)$.)

\subsection{Interpretation in terms of deformed worldsheet theory}

Let us recapitulate what we have done up to now. We have determined the $4$-point functions of the dual CFT from the worldsheet, see Section~\ref{sec:4.3}. Using these $4$-point functions we have then computed the anomalous conformal dimensions of the symmetric orbifold fields, replicating essentially the analysis of  \cite{Gaberdiel:2015uca,Apolo:2022fya}. The fact that the anomalous conformal dimensions have come out correctly thus confirms that not only the general structure of the symmetric orbifold amplitudes can be reproduced from the worldsheet \cite{Eberhardt:2019ywk}, but also their relative normalisations, etc., see in particular Section~\ref{sec:normalisation}. This is a non-trivial consistency check that goes beyond what had been checked before. 

However, there should also be a more direct worldsheet description of the perturbed system. Indeed, the perturbed symmetric orbifold theory should still have a string theory dual, and we should be able to give an explicit description of the corresponding worldsheet theory. In fact, our analysis above essentially shows what the relevant worldsheet theory is: it is simply obtained by perturbing the original worldsheet theory by the exactly marginal operator $\int d^2x \, \bigl( \tilde{G}_{-1} \bar{\tilde{G}}_{-1} {\bf \Phi}\bigr)(z,x)$, i.e.\ by adding the term, see also \cite{Giveon:2017nie} for a similar construction in another context, 
\be\label{pert}
\int d^2z \int d^2x \, \bigl(\tilde{G}_{-1} \bar{\tilde{G}}_{-1} {\bf \Phi} \bigr)(z,x) 
\ee
to the worldsheet action. Indeed, the above calculations show that $\int d^2z \, \bigl(\tilde{G}_{-1} \bar{\tilde{G}}_{-1}{\bf \Phi}\bigr)(z,x)$ inserts the perturbing field $\Phi(x)$ in the dual CFT correlator, and the integral over $x$ then amounts to performing the dual CFT perturbation. Note that in calculating the worldsheet correlators there is some freedom in where the $\tilde{G}_{-1}$  (and $\bar{\tilde{G}}_{-1}$) operators should be inserted; for the calculations above it was convenient to let them act on the currents, but the result would have been unchanged if they had been applied to the perturbing fields. In fact, in the general case, this is more natural since each additional perturbing fields comes with an additional $\tilde{G}_{-1}\bar{\tilde{G}}_{-1}$ insertion, i.e.\ should be described by (\ref{pert}).\footnote{More explicitly, the number of $\tilde{G}^-_{-1}$ operators equals $(n+\ell-2+2g)$ if we perturb the $n$-point function to $\ell$'th order, see \cite[eq.~(3.8)]{Dei:2020zui}. Since among the original $n$ fields we have already distributed $(n-2+2g)$ $\tilde{G}^-_{-1}$ operators (so that it makes sense before perturbation), each of the additional $\ell$ perturbing fields must come with its own $\tilde{G}^-_{-1}$ descendant, and similarly for the right-movers. We should also mention that the appropriate number of (vacuum) $B$ fields will have to be added in order to neutralise the overall $U_0$ charge.} The presence of these $\tilde{G}_{-1}\bar{\tilde{G}}_{-1}$ descendants also guarantees that the integrand has conformal dimension $(1,1)$ on the worldsheet, so that the perturbation term is marginal. Since the spacetime field $\Phi$ also has spacetime conformal dimension $(1,1)$, likewise the $x$ integral does not break the conformal symmetry. Incidentally, because of the localisation properties of the correlators, there is in fact only one integral to perform since the delta functions will effectively identify $x$ and $z$. 

\section{Conclusion}\label{con}

In this paper we have analysed the perturbation of the symmetric orbifold theory by the exactly marginal operator coming from the $2$-twisted sector using the dual worldsheet perspective. In particular, we have identified how the worldsheet theory needs to be perturbed, see eq.~(\ref{pert}), in order to continue to be dual to the perturbed spacetime theory. We have checked by explicit calculations that the anomalous conformal dimensions of some simple spacetime currents are correctly reproduced by this worldsheet prescription. 

Given the explicit worldsheet description of the perturbing operator, it should now be possible to study various structural properties of the perturbed theory from the worldsheet perspective. For example, for the dual perturbed CFT there is a relatively simple criterion for when an operator picks up an anomalous dimension, see e.g.\ \cite{Cardy:1989da,Fredenhagen:2007rx,Gaberdiel:2013jpa}, and it would be interesting to see if there is a corresponding condition that can be formulated directly in terms of a worldsheet OPE. Similarly, it would be interesting to understand what condition guarantees that the worldsheet deformation is exactly marginal in the dual CFT. The AdS$_3$ case furnishes an interesting arena in which many of these questions can be studied in detail, but ultimately we would also like to apply these ideas to the case of AdS$_5$, i.e.\ to the worldsheet theory proposed in \cite{Gaberdiel:2021iil,Gaberdiel:2021jrv}. We hope to return to some of these questions elsewhere.

\acknowledgments
We thank Bob Knighton, Rajesh Gopakumar, Beat Nairz, and Jakub Vo\v{s}mera for useful discussions and comments on a draft of this paper. The work of MAF, KN and VS was supported by a grant from the Swiss National Science Foundation. The activity of the whole group is generally supported by the NCCR SwissMAP that is also supported by the Swiss National Science Foundation. Finally, MRG acknowledges support from the Simons Foundation grant 994306 (Simons Collaboration on Confinement and QCD Strings).

\appendix

\section{$\mathcal{N}=4$ superconformal algebra in the $\mathbb{T}^4$ CFT}
\label{app:N=4ofT4}
In this appendix, we explain our conventions for the CFT on $\mathbb{T}^4$, both for the space-time theory and the worldsheet theory (where we consider the topologically twisted version). Consider one copy of the (holomorphic part of the) $\mathbb{T}^4$ CFT described by two complex chiral bosons $X^{1}$, $X^{2}$ (as well as their complex conjugates  $\bar{X}^{1}$, $\bar{X}^{2}$) and two complex chiral fermions $\psi^{1}$, $\psi^{2}$, again together with their complex conjugates $\bar{\psi}^{1}$, $\bar{\psi}^{2}$. They satisfy
\begin{equation}
\partial X^{a}(z) \, \partial \bar{X}^{b}(0)
\sim \frac{\delta^{ab}}{z^2} \ ,
\qquad \qquad
\psi^{a}(z) \, \bar{\psi}^{b}(0)
\sim \frac{\delta^{ab}}{z} \ .
\end{equation}
The small $\mathcal{N}=4$ superconformal symmetry of this theory ($c=6$) is described by, see e.g.~\cite{Brunner:2006tc}:
\begin{equation}
\begin{aligned}
T= \partial X^{a} \partial \bar{X}^{a}
&+
\frac{1}{2}\left(\partial \psi^{a}\bar{\psi}^{a}+\partial \bar{\psi}^{a}\psi^{a}\right) \ ,
\\
G^+= \partial\bar{X}^{a}\psi^{a} \ ,
\qquad&\qquad
\tilde{G}^+ = -\partial X^1 \psi^2 + \partial X^2 \psi^1 \ ,
\\
G^-= \partial X^{a}\bar{\psi}^{a} \ ,
\qquad&\qquad
\tilde{G}^- = -\partial \bar{X}^1 \bar{\psi}^2 + \partial \bar{X}^2 \bar{\psi}^1 \ ,
\\
J=\mathcal{J}^3= \frac{1}{2}\psi^{a}\bar{\psi}^{a} \ ,
\qquad\qquad
&\mathcal{J}^+=\psi^{1}\psi^{2} \ ,
\qquad\qquad
\mathcal{J}^-=-\bar{\psi}^{1}\bar{\psi}^{2} \ ,
\end{aligned}
\end{equation}
where the repeated indices are summed over $a\in\{1,2\}$. $T$, $G^\pm$ and $J$ generate the $\mathcal{N}=2$ superconformal algebra, see e.g.\
\cite{Gaberdiel:2021njm}. In particular in our conventions, the eigenvalues of $G^{\pm}$ and $\tilde{G}^{\pm}$ under $J$ are $\pm \frac{1}{2}$ and
\begin{equation}
G^+(z) G^-(0) \sim \frac{c/3}{z^3}+\frac{2J(0)}{z^2}+\frac{T+\partial J}{z} \ ,
\quad J(z) J(0) \sim \frac{c/12}{z^2} \ .
\end{equation}
$\mathcal{J}^3$, $\mathcal{J}^\pm$ generate the  $\mathfrak{su}(2)_1$ R-symmetry, normalised as
\begin{equation}
\mathcal{J}^3(z) \mathcal{J}^3(0)
\sim
\frac{1/2}{z^2} \ , \qquad
\mathcal{J}^3(z) \mathcal{J}^\pm (0)
\sim
\frac{\pm \mathcal{J}^\pm}{z} \ , \qquad
\mathcal{J}^+(z)\mathcal{J}^-(0)
\sim
\frac{1}{z^2}+\frac{2\mathcal{J}^3}{z} \ .
\end{equation}

In the worldsheet description of string theory on $\text{AdS}_3 \times \text{S}^3 \times \mathbb{T}^4$, the topologically twisted version of the $\mathcal{N}=4$ of $\mathbb{T}^4$ appears, which we differentiate by the subscript $\mathbb{T}^4$ if necessary. This means that we set
\begin{equation}
T_{\mathbb{T}^4} = T + \partial J \ .
\end{equation}
In this convention, $G^+_{\mathbb{T}^4}$ and $G^-_{\mathbb{T}^4}$ have weights $1$ and $2$ respectively, while $T_{\mathbb{T}^4}$ is primary of weight $2$. We also have
\begin{equation}
G^+_{\mathbb{T}^4}(z) G^-_{\mathbb{T}^4}(0) \sim \frac{c/3}{z^3}+\frac{2J_{\mathbb{T}^4}(0)}{z^2}+\frac{T_{\mathbb{T}^4}}{z} \ , \qquad
T_{\mathbb{T}^4}(z) J_{\mathbb{T}^4}(0) \sim \frac{-c/6}{z^3}+\frac{J_{\mathbb{T}^4}(0)}{z^2}+\frac{\partial J_{\mathbb{T}^4}}{z} \ .
\end{equation}

In the hybrid formalism, it is convenient to bosonise the fermions of $\mathbb{T}^4$. In particular, we set \cite{Berkovits:1999im}
\begin{equation}
\psi^a = e^{iH^a_{\mathbb{T}^4}} \ , \qquad\qquad \bar{\psi}^a = e^{-iH^a_{\mathbb{T}^4}} \ ,
\end{equation}
where $H^a_{\mathbb{T}^4}$ are bosons satisfying
\begin{equation}
H^a_{\mathbb{T}^4}(z) H^b_{\mathbb{T}^4}(0) \sim -\delta^{ab} \ln(z) \ .
\end{equation}
The topologically twisted stress-tensor equals
\begin{equation} \label{eq:t-of-t4}
T_{\mathbb{T}^4} = \sum_a \left(\partial X^{a} \partial \bar{X}^{a} - \frac{1}{2} (\partial H^a_{\mathbb{T}^4})^2 + \frac{1}{2} \partial^2(i H^a_{\mathbb{T}^4})\right) \ .
\end{equation}
This means that the bosons $H^a_{\mathbb{T}^4}$ have background charges $\Lambda_{H^a_{\mathbb{T}^4}}=1$, see \cite{Blumenhagen:2013fgp}.

\section{Free field realisation of $\mathfrak{psu}(1,1|2)_1$} \label{app:psu1}

In this appendix we review the free field realisation of $\mathfrak{psu}(1,1|2)_1$ following \cite{Dei:2020zui}. The relevant free fields comprise $4$ symplectic bosons ($\xi^{\pm}$ and $\eta^{\pm}$) and $4$ free fermions ($\chi^{\pm}$ and $\psi^{\pm}$) \cite{Dei:2020zui,Eberhardt:2018ouy, Lesage:2002ch, Gotz:2006qp, Ridout:2010jk, Quella:2013oda}, with (anti)commutation relations
\begin{equation} \label{eq:ff_commutators}
\{\psi^\alpha_r, \chi^\beta_s\} = \epsilon^{\alpha\beta} \delta_{r+s,0} \ , \qquad\qquad
[\xi^\alpha_r, \eta^\beta_s] = \epsilon^{\alpha\beta} \delta_{r+s,0} \ , 
\end{equation}
where $\alpha, \beta \in \{\pm\}$, $\epsilon^{+-} = 1 = -\epsilon^{-+}$ and the other combinations vanish. They generate the algebra $\mathfrak{u}(1,1|2)_1$ via
\begin{equation} \label{eq:u-currents}
\begin{split}
U &= -\tfrac{1}{2}\eta^+\xi^-+\tfrac{1}{2}\eta^-\xi^+ \ , \\
J^3 &= -\tfrac{1}{2}\eta^+\xi^--\tfrac{1}{2}\eta^-\xi^+ \ , \\
J^\pm &= \eta^\pm\xi^\pm  \ , \\
S^{\alpha\beta+} &=\chi^\beta\xi^\alpha \ ,
\end{split}
\qquad\qquad
\begin{split}
V &= -\tfrac{1}{2}\chi^+\psi^-+\tfrac{1}{2}\chi^-\psi^+ \ , \\
K^3 &= -\tfrac{1}{2}\chi^+\psi^--\tfrac{1}{2}\chi^-\psi^+ \ , \\
K^\pm &= \pm \chi^\pm \psi^\pm \ , \\
S^{\alpha\beta-} &= \eta^\alpha\psi^\beta \ ,
\end{split}
\end{equation}
see also \cite{Gaberdiel:2022als}. Note that relative to \cite{Dei:2020zui} we have removed a minus sign in front of $S^{\alpha\beta-}$. Here $J^a$ and $K^a$ generate $\mathfrak{sl}(2,\mathbb{R})_1$ and $\mathfrak{su}(2)_1$, respectively, while $S^{\alpha\beta\gamma}$ correspond to the odd elements of the superalgebra. In order to reduce this to $\mathfrak{psu}(1,1|2)_1$, one has to set $Z=U + V = 0$; this can be achieved by considering those states that are annihilated by $Z_n$ with $n\geq 0$, since then the $Z_{-n}$ descendants are null. 

The representations that are relevant for us are the spectrally flowed images of the NS- and R-sector representation. In the R-sector, the symplectic bosons are integer moded, and their zero modes are taken to act on the ground states (that are annihilated by the positive modes) as 
\begin{equation} \label{eq:Rgs}
\begin{split}
\xi^-_0 \ket{m_1, m_2} &= -\ket{m_1 - \tfrac{1}{2}, m_2} \ , \\
\xi^+_0 \ket{m_1, m_2} &= \ket{m_1, m_2 + \tfrac{1}{2}} \ , \\
\psi^+_0 \ket{m_1, m_2} &= 0 \ ,
\end{split}
\qquad\qquad
\begin{split}
\eta^+_0 \ket{m_1, m_2} &= 2m_1\ket{m_1 + \tfrac{1}{2}, m_2} \ , \\
\eta^-_0 \ket{m_1, m_2} &= -2m_2\ket{m_1, m_2 - \tfrac{1}{2}} \ , \\
\chi^+_0 \ket{m_1, m_2} &= 0 \ .
\end{split}
\end{equation}
Here $m_1$ and $m_2$ label the charge with respect to the two conjugate symplectic bosons, i.e.\ with respect to $\mp \eta^\pm\xi^\mp = U \pm J^3$,
\begin{equation}
\begin{aligned}
-(\eta^+\xi^-)_0 \ket{m_1, m_2} &= +(2m_1-\tfrac{1}{2}) \ket{m_1, m_2} \ , \\
+(\eta^-\xi^+)_0 \ket{m_1, m_2} &= -(2m_2+\tfrac{1}{2}) \ket{m_1, m_2} \ .
\end{aligned}
\end{equation}
In the NS-sector (that will only play a minor role in the paper), the fields are half-integer moded, and the ground state $\ket{0}$ is simply characterised by 
\begin{equation}
\xi^\pm_r\ket{0}=\eta^\pm_r\ket{0}=\psi^\pm_r\ket{0}=\chi^\pm_r\ket{0}=0 \qquad \text{for }r\geq \tfrac{1}{2} \ .
\end{equation}

We do not just need these highest weight representations, but also their spectrally flowed images: for any automorphism $\tau$ of the free field algebra, we define the $\tau$-spectrally flowed representation ${\cal H}^{\tau}$ of ${\cal H}$ (where ${\cal H}$ is any of the above highest weight representations) by twisting the action of the modes on ${\cal H}$. This is to say,  the action of the modes is defined on $ [\ket{v}]^\tau\in {\cal H}^\tau$ via 
\begin{equation}
O_n ~ [\ket{v}]^\tau  = [\tau(O_n) ~ \ket{v}]^\tau \ ,
\end{equation}
where $O_n$ is an element of the algebra. Our free field algebra has two natural spectral flow automorphisms, $\sigma^{(\pm)}$, that are defined by (the other actions are all trivial)
\begin{equation} \label{eq:spectral_flows}
\begin{split}
\sigma^{(+)}(\eta^+_r) &= \eta^+_{r-\frac{1}{2}} \ , \\
\sigma^{(+)}(\xi^-_r) &= \xi^-_{r+\frac{1}{2}} \ , \\
\sigma^{(+)}(\chi^+_r) &= \chi^+_{r+\frac{1}{2}} \ , \\
\sigma^{(+)}(\psi^-_r) &= \psi^-_{r-\frac{1}{2}} \ , \\
\end{split}
\qquad\qquad
\begin{split}
\sigma^{(-)}(\xi^+_r) &= \xi^+_{r-\frac{1}{2}} \ , \\
\sigma^{(-)}(\eta^-_r) &= \eta^-_{r+\frac{1}{2}} \ , \\
\sigma^{(-)}(\psi^+_r) &= \psi^+_{r+\frac{1}{2}} \ , \\
\sigma^{(-)}(\chi^-_r) &= \chi^-_{r-\frac{1}{2}} \ .\\
\end{split}
\end{equation}
It is convenient to combine them as 
\begin{equation}
\sigma = \sigma^{(+)} \circ \sigma^{(-)} \ , \qquad\qquad
\hat{\sigma} = \sigma^{(+)} \circ (\sigma^{(-)})^{-1} \ ,
\end{equation}
since $\sigma$ is then the usual spectral flow automorphism of $\mathfrak{psu}(1,1|2)_1$.
The $\hat{\sigma}$ spectral flow plays an important role for the construction of a $\mathfrak{psu}(1,1|2)_1$ vacuum $B(z)$ with non-zero eigenvalues $U_0=1$ and $V_0=-1$ --- this field was called $W(z)$ in \cite{Dei:2020zui}, see in particular eq.\ (2.30) of that paper
\begin{equation}
B(z)=\text{V}([\psi^+_{-\frac{3}{2}}\psi^-_{-\frac{3}{2}}\psi^+_{-\frac{1}{2}}\psi^-_{-\frac{1}{2}}\ket{0}]^{\hat{\sigma}^2},z) \ ,
\end{equation}
where $\text{V}(\phi,z)$ is the usual CFT vertex operator associated to the state $\phi$, see e.g.\ \cite{meromorphic}. We further introduce the field $\tilde{B}(z)$
\begin{equation}
\tilde{B}(z)=\text{V}([\psi^+_{-\frac{1}{2}}\psi^-_{-\frac{1}{2}}\ket{0}]^{\hat{\sigma}^2},z) \ ,
\qquad\qquad
B=(\psi^+\psi^-\tilde{B}) 
\label{A11}
\end{equation}
that is now uncharged with respect to the fermions, i.e.\ satisfies $V_0=0$ but still has $U_0=1$. This rewriting is convenient for decoupling the fermions from the bosonic part of the correlators.

Finally, for any worldsheet vertex operator $V(\phi,z)$, the spacetime $x$-dependence is introduced by conjugation with the translation operator in the dual CFT \cite{Eberhardt:2019ywk}
\be
\text{V}(\phi,z,x) = e^{xJ^+_0} \text{V}(\phi,z,0)e^{-xJ^+_0} \ .
\ee
For the vertex operators associated to the states $[\ket{m_1, m_2}]^{\sigma^w}$ we shall use the short-hand notation 
\begin{equation} \label{eq:tw_ground_states}
V^w_{m_1, m_2}(z,x) = \text{V}([\ket{m_1, m_2}]^{\sigma^w}, z,x) \ .
\end{equation}

\section{Review of the free field realisation of the hybrid formalism} \label{app:hybrid}
In this appendix, we review the hybrid formalism with pure NS-NS flux $k=1$, written in terms of the free field realisation of $\mathfrak{u}(1,1|2)_1$ \cite{Berkovits:1999im,Gaberdiel:2022als,gerick:thesis}. The worldsheet theory
consists of three parts: 
\begin{enumerate}
\item a $\mathfrak{psu}(1,1|2)_1$ WZW model,
\item ghosts $\rho$ and $\sigma$,
\item a topologically twisted $\mathbb{T}^4$.
\end{enumerate}
Appendix~\ref{app:psu1} reviews the free field realisation of $\mathfrak{psu}(1,1|2)_1$. The bosons $\rho$ and $\sigma$ satisfy
\begin{equation}
\rho(z) \rho(\zeta) \sim -\ln(z-\zeta) \ , \qquad \sigma(z) \sigma(\zeta) \sim -\ln(z-\zeta) \ .
\end{equation}
$\sigma$ is the boson that appears in the bosonisation of the $bc$ diffeomorphism ghost with central charge $c=-26$. The boson $\rho$ is a combination of different fields, see \cite{Berkovits:1999im}. Appendix~\ref{app:N=4ofT4} explains our conventions for the topologically twisted $\mathbb{T}^4$.

As we mention in Section~\ref{sec:states}, see also Appendix~\ref{app:psu1}, the free field realisation of $\mathfrak{psu}(1,1|2)_1$ is obtained from the one of $\mathfrak{u}(1,1|2)_1$ once we effectively set $Z=0$, where $Z=U+V$. This can be done by considering the subspace that is annihilated by $Z_n$ for $n\geq 0$ \cite{Dei:2020zui}.
It is only on this subspace that the supercurrents in eq.~(\ref{eq:u-currents}) written as bilinears of the free fields give the correct $\mathfrak{psu}(1,1|2)_1$ anti-commutation relations.
Similarly, the hybrid $\mathcal{N}=4$ currents written naively in terms of the free fields lead to spurious terms proportional to $Z$. In order to write down an exact $\mathcal{N}=4$ algebra, additional ghosts have been introduced in \cite{Gaberdiel:2022als}:
\begin{equation}
b(z) c(\zeta) \sim \frac{1}{z-\zeta} \ , \qquad b^{\prime}(z) c^{\prime}(\zeta)\sim \frac{1}{z-\zeta} \ , \qquad \beta^{\prime}(z) \gamma^{\prime}(\zeta) \sim \frac{1}{z-\zeta} \ ,
\end{equation} 
where $(b,c)$ and $(b^{\prime},c^{\prime})$ are anti-commuting ghosts with weights $(1,0)$, and $(\beta^{\prime},\gamma^{\prime})$ are commuting ghosts with weights $(1,0)$ \cite{Blumenhagen:2013fgp}. In terms of these, the twisted $\mathcal{N}=4$ generators with $c=6$ are then \cite{Berkovits:1999im,Gaberdiel:2022als,Naderi:2022bus}
\begin{subequations}
\begin{align}
T&=T_{\text{int}}-\frac{1}{2} (\partial \rho)^2-\frac{1}{2} (\partial \sigma)^2+\frac{3}{2} \partial^2(\rho+i\sigma)+T_{\mathbb{T}^4} \ , \label{eq:t-hybrid} \\
G^+&=e^{-\rho} Q + e^{i\sigma} T - \partial(e^{i\sigma}\partial(\rho+iH_{\mathbb{T}^4}))+G^+_{\mathbb{T}^4} \\
G^-&=e^{-i\sigma} \\
J&=\frac{1}{2}\partial(\rho+i\sigma+iH_{\mathbb{T}^4}) \ , \\
J^{\pm\pm}&=e^{\pm(\rho+i\sigma+iH_{\mathbb{T}^4})} \ , \\
\tilde{G}^+&=e^{\rho+iH_{\mathbb{T}^4}} \ , \\
\tilde{G}^-&=e^{-2\rho-i\sigma-iH_{\mathbb{T}^4}} Q - e^{-\rho-iH_{\mathbb{T}^4}} T - e^{-\rho-i\sigma} \tilde{G}^-_{\mathbb{T}^4} + e^{-\rho-i\sigma} [i\partial \sigma \partial(\rho+iH_{\mathbb{T}^4})+\partial^2(\rho+iH_{\mathbb{T}^4})] \label{eq:gtilde-}\ ,
\end{align}
\end{subequations}
where the fields with subscript $\mathbb{T}^4$ are defined in Appendix~\ref{app:N=4ofT4}, and 
\begin{equation}
\begin{aligned}
Q &= (\chi^+ \chi^-) (\xi^+ \partial \xi^- - \xi^- \partial \xi^+) \ , \\
T_{\text{int}}&=T_{\mathfrak{u}(1,1|2)_1}+T_{bc}+T_{b^{\prime}c^{\prime}}+T_{\beta^{\prime}\gamma^{\prime}}+\partial(\gamma^{\prime} Z) \ , \\
T_{bc}&=((\partial c)b) \ , \qquad T_{b^{\prime}c^{\prime}}=((\partial c^{\prime})b^{\prime}) \ , \qquad T_{\beta^{\prime} \gamma^{\prime}}=((\partial \gamma^{\prime})\beta^{\prime}) \ .
\end{aligned}
\end{equation}
Note that, as explained in \cite{Gaberdiel:2021njm,Naderi:2022bus}, we have applied the similarity transformation $R$ defined below to the fields in \cite{Gaberdiel:2022als},
\begin{equation}
R = \oint e^{i\sigma} G^-_{\mathbb{T}^4} \ .
\end{equation}
This simplifies the form of the physical states \cite{Gaberdiel:2021njm}. Moreover, we follow the normal-ordering prescription explained in \cite{Gaberdiel:2022als}.

The picture number is defined as the eigenvalue of the zero mode of
\begin{equation}
P=-\partial(\phi+i\kappa) \ ,
\end{equation}
where $\phi$ and $\kappa$ appear in the bosonisation of the $\beta\gamma$ system \cite{Blumenhagen:2013fgp}.
In particular, any field of the form 
\begin{equation}
e^{m\rho+in \sigma+ik_1 H^1_{\mathbb{T}^4}+ik_2 H^2_{\mathbb{T}^4}} D
\quad
\text{has picture number }(-m)
\ ,
\end{equation}
where $D$ is a combination of various fields which does not involve any exponentials of $\rho$, $\sigma$ and $H^a_{\mathbb{T}^4}$.
There is finally a picture raising operator $P_+$ defined as \cite{Blumenhagen:2013fgp}
\begin{equation}
P_+ = - G^+_0 (e^{-\rho-iH_{\mathbb{T}^4}})_0 \ ,
\end{equation}
which maps a physical state to a physical state, and increases the picture number.

\subsection{Bosonising the fermions}\label{app:C1}

It will also often be convenient to bosonise the fermions $\chi^{\pm}$ and $\psi^{\pm}$ of $\mathfrak{u}(1,1|2)_1$ by introducing two bosons $q_i$ with $i\in\{1,2\}$ such that
\begin{equation}
q_i(z) q_j(\zeta) \sim -\delta_{ij}\, \ln(z-\zeta) \ .
\end{equation}
Then the fermions can be written as 
\begin{equation} \label{eq:ferm-bosonisation}
\chi^+=e^{iq_1} \ , \qquad \psi^-=-e^{-iq_1} \ , \qquad \psi^+ = e^{iq_2} \ , \qquad \chi^-=e^{-iq_2} \ ,
\end{equation}
and it is easy to verify that this reproduces eq.~(\ref{eq:opes-free fields}). Here we have not written out explicitly the cocycle factors \cite{Blumenhagen:2013fgp,Goddard:1986ts,Burrington:2015mfa} that guarantee that the fields satisfy fermionic statistics, i.e.\ 
\begin{equation} \label{eq:cocycle-fermions}
e^{i k_2 q_2(z)} e^{i k_1 q_1(w)} = (-1)^{k_1 k_2} e^{i k_1 q_1(w)} e^{i k_2 q_2(z)} \ .
\end{equation}

The correlation functions of the free boson theory can be calculated straightforwardly. For a boson with background charge $\Lambda$ and OPE 
\be
X(z) X(w) \sim - c_X \ln(z-w) \ ,
\end{equation}
the correlation function of the exponentials takes the form \cite{Blumenhagen:2013fgp,Polchinski:1998rr}
\begin{equation} \label{eq:cor-exp}
\Bigl\langle  \prod_{j=1}^{n} e^{ik_j X(z_j)}\Bigr\rangle = \delta\Bigl(\sum_{j=1}^n k_j - \Lambda_X\Bigr)\,  \prod_{l>j} (z_j-z_l)^{c_X k_j k_l}\ .
\end{equation}
From this, one can then also deduce the correlation functions of for example $\partial X$, by taking suitable derivatives.

\section{The symplectic boson correlators}\label{WW}

In this appendix we explain how the symplectic boson correlators can be evaluated. 

\subsection{The $\xi^{\pm}$ Ward identities}
We begin by defining 
\begin{equation} \label{eq:Ppm}
{\cal P}^\pm(z)
=\frac{\prod_{i=1}^4 (z-z_i)^{\frac{w_i+1}{2}}}{(z-z_5) (z-z_6)} \, \Big\langle \xi^\pm(z) \, B(z_6) B(z_5) \, 
\prod_{i=1}^4 V^{w_i}_{m_1^i, m_2^i}(z_i,x_i)\Big\rangle \ ,
\end{equation}
where $B$ is defined in eq.~\eqref{A11} and where the $V$s are vertex operators of spectrally-flowed Ramond ground states, see eq.~\eqref{eq:tw_ground_states}. We are primarily interested in the case where $w_1=w_2=2$ and $w_3=w_4=1$. The prefactor 
is chosen so as to cancel all poles and zeros from the correlator as $z$ approaches the points $z_i$ with $i=1,2,\ldots,6$.  The only remaining pole is thus at $z\rightarrow\infty$, where $P^\pm(z) $ goes as $z^N$ with
\begin{equation}
N=1+\sum_{i=1}^n\frac{w_i-1}{2} \ .
\end{equation}
Liouville's theorem then implies that the $z$-dependence is a polynomial of degree $N$,
\begin{equation}
{\cal P}^\pm(z)=\sum_{k=0}^N P^\pm_k z^k \ .
\end{equation}
In our situation, $N=2$ so we have altogether six unknown coefficients, three from ${\cal P}^+(z)$ and three from ${\cal P}^-(z)$.

By construction, the functions ${\cal P}^\pm(z)$ satisfy the ``incidence relation'' \cite{Dei:2020zui}
\be
\left({\cal P}^-+x_i {\cal P}^+\right)^{(\ell)}(z_i) = 0 \ , \qquad\qquad 0\leq\ell\leq w_i-1 \ . \label{eq:IncidenceRelation}
\ee
In terms of the six coefficients $P^\pm_k$ of ${\cal P}^\pm(z)$, eq.~(\ref{eq:IncidenceRelation}) is a homogeneous linear system.
A necessary condition for a non-trivial solution is therefore that the determinant of this linear system vanishes, and this gives rise to the delta function localisation.\footnote{In fact, the determinant contains at most a term linear in $x_4$, as can be deduced from the structure of the matrix, and hence must be of the form $(x_4-\Gamma(z_4))\times h(x_{i\neq4},z_i)$ which precisely implies the delta function dependence.} Provided that a non-trivial solution exists, all unknowns can be solved in terms of (and are proportional to) $P^+_0$ for example. In fact, the ratio 
\begin{equation}\label{covering}
\Gamma(z) = -\frac{{\cal P}^-(z)}{{\cal P}^+(z)}
\end{equation}
is fully determined and it defines precisely the relevant covering map since it satisfies \cite{Dei:2020zui}
\begin{equation} \label{eq:CoveringDefinition}
\Gamma(z)\sim x_i + \frac{1}{w_i!}\Gamma^{(w_i)}(z_i)(z-z_i)^{w_i}+\cdots \ , \qquad\qquad z\rightarrow z_i \ .
\end{equation}

It is sometimes convenient to remove only the pole at $z=z_i$, i.e.\ to consider instead of $P^\pm(z)$, defined by eq.~(\ref{eq:Ppm}), the functions
\be\label{PPi}
{\cal P}_i^\pm(z) = {\cal P}_i(z) \, {\cal P}^\pm(z) = (z-z_i)^{\frac{w_i+1}{2}} \, \, \Big\langle \xi^\pm(z) \, B(z_6) B(z_5) \, 
\prod_{i=1}^4 V^{w_i}_{m_1^i, m_2^i}(z_i,x_i)\Big\rangle  \ , 
\ee
where ${\cal P}_i(z)$ is simply 
\be\label{Pi}
{\cal P}_i(z) = \frac{(z-z_5)(z-z_6)}{\prod_{j\neq i=1}^{4} (z-z_j)^{(w_j+1)/2}} \ . 
\ee
The functions ${\cal P}_i^\pm(z)$ are still analytic in a neighbourhood of $z=z_i$ and thus can be expressed in terms of their Taylor series as 
\begin{equation}
{\cal P}_i^\pm(z) =\sum_{\ell=0}^\infty \frac{1}{\ell!}\big({\cal P}_i^\pm\big)^{(\ell)}(z_i)\, (z-z_i)^\ell \ . 
\end{equation}
Conversely, ${\cal P}_i^\pm(z)$ can also be determined by evaluating the OPE of $\xi^\pm(z)$ near $z_i$, and this leads to 
\be
\begin{aligned}
{\cal P}_i^+(z) & =\sum_{\ell=0}^\infty \frac{1}{\ell!}\big({\cal P}_i^+\big)^{(\ell)}(z_i)\, (z-z_i)^\ell 
= \sum_{\ell=0}^\infty \langle[\xi^+_{-\ell}\rangle_i (z-z_i)^\ell \\
{\cal P}_i^-(z) & =\sum_{\ell=0}^\infty \frac{1}{\ell!}\big({\cal P}_i^-\big)^{(\ell)}(z_i)\, (z-z_i)^\ell 
= \sum_{\ell=0}^\infty \langle[\xi^-_{-\ell}\rangle_i (z-z_i)^{\ell+w_i}
-x_i \sum_{\ell=0}^\infty \langle[\xi^+_{-\ell}\rangle_i (z-z_i)^\ell \ ,\label{B.8}
\end{aligned}
\ee
where $\langle[\xi^\pm_{-\ell}\rangle_i$ is the correlator with the mode $\xi^\pm_{-\ell}$ acting \emph{inside} the spectral flow bracket at the $i$-th insertion, e.g.\ $\langle[\xi^\pm_{-\ell}\rangle_1$ denotes the correlator.
\begin{equation} \label{eq:short_inside_modes}
\langle[\xi^\pm_{-\ell}\rangle_1=\Big\langle
B(z_6) B(z_5)\, 
\text{V}\left([\xi^\pm_{-\ell}|m_1^1, m_2^1\rangle]^{\sigma^{w_1}},z_1,x_1\right)\prod_{i=2}^4 V^{w_i}_{m_1^i, m_2^i}(z_i,x_i)\Big\rangle \ .
\end{equation}
Comparing coefficients in eq.~(\ref{B.8}), we therefore have for $\ell\geq 0$,
\begin{equation}
\langle[\xi^+_{-\ell}\rangle_i   = \frac{1}{\ell!}\big({\cal P}_i^+\big)^{(\ell)}(z_i)\ ,
\qquad\qquad
\langle[\xi^-_{-\ell}\rangle_i   =  \frac{1}{\ell!}\big({\cal P}_i^- + x_i {\cal P}_i^+\big)^{(\ell+w_i)}(z_i)\ .
\end{equation}
In particular, $\langle[\xi^+_0\rangle_i = {\cal P}_i^+(z_i)$, and
\begin{equation} \label{B.23} 
\begin{aligned}
\langle[\xi^-_0\rangle_i & = \frac{1}{w_i!}
\frac{d^{w_i}}{dz_i^{w_i}} ({\cal P}_i^-+ x_i {\cal P}_i^+)(z_i) 
= -   \frac{1}{w_i!}
\frac{d^{w_i}}{dz^{w_i}}\Big( {\cal P}_i(z) {\cal P}^+(z)\big(\Gamma(z)-x_i\big)\Big)\Big\vert_{z=z_i} \\
& =-{\cal P}_i^+(z_i) \, a_i \ , 
\end{aligned}
\end{equation}
where we have used eqs.~(\ref{PPi}), (\ref{covering}), as well as (\ref{eq:CoveringDefinition}), and where $a_i$ is defined as 
\begin{equation}
a_i=\left. \frac{1}{w_i!} \frac{d^{w_i}}{dz^{w_i}} \Gamma(z)\right|_{z=z_i}\ .
\end{equation}
On the other hand, because of eq.~(\ref{eq:Rgs}), we know that the effect of zero-modes is just to translate the values of $m_1^i$ and $m_2^i$, which we write as
\begin{equation}
\langle[\xi^+_0\rangle_i = \langle m^i_2+\tfrac{1}{2}\rangle \ , \qquad
\langle[\xi^-_0\rangle_i = -\langle m_1^i-\tfrac{1}{2}\rangle \ ,
\end{equation}
(cf.\ eq.~\eqref{eq:short_inside_modes}) and thus we obtain the identities 
\begin{equation} \label{eq:Recurrence}
\langle m^i_2+\tfrac{1}{2}\rangle={\cal P}_i^+(z_i) \ , \qquad
\langle m^i_1-\tfrac{1}{2}\rangle= {\cal P}_i^+(z_i) \, a_i  = \langle m^i_2+\tfrac{1}{2}\rangle\, a_i\ .
\end{equation}

The second equation in eq.~\eqref{eq:Recurrence} is the recurrence relation (4.28) of \cite{Dei:2020zui}. Since on the symplectic boson ground states the eigenvalues  of $J^3_0$ and $U_0$ are, see eqs.~(\ref{eq:u-currents}) and (\ref{eq:Rgs}),
\be
j_i = m^i_1+m^i_2+\frac{w_i}{2} \ , \qquad\qquad u_i = m^i_1 - m^i_2-\frac{1}{2} \ ,
\ee
the recurrence relation implies that 
\begin{equation}\label{shift}
\langle j_i+1, u_i \rangle=\frac{1}{a_i}\, \langle j_i, u_i\rangle \ ,
\end{equation}
and hence the dependence of the correlator on $m^i_1+m^i_2$ has the form\footnote{In principle the right-hand side could be multiplied by a function that is periodic in $j_i \mapsto j_i+1$. However, as in \cite{Dei:2020zui}, we assume that this periodic function is trivial.}
\begin{align}
\langle B(z_5)B(z_6)&V^{w_1}_{m^1_1, m^1_2}(z_1,x_1)V^{w_2}_{m^2_1, m^2_2}(z_2,x_2)V^{w_3}_{m^3_1, m^3_2}(z_3,x_3)V^{w_4}_{m^4_1, m^4_2}(z_4,x_4)\rangle \nonumber \\
&\quad = C(u_l;x_l,z_l)\, \prod_{i=1}^4 a_i^{-(m_1^i+m_2^i)} \ . \label{eq:SolutionC} 
\end{align}
In order to determine the dependence of $C(u_l;x_l,z_l)$ on the $u_l$ we note that the first equation in eq.~\eqref{eq:Recurrence} implies that 
\be
b_i \, \langle m^i_2+\tfrac{1}{2}\rangle =  b_j \langle m^j_2+\tfrac{1}{2}\rangle \ , 
\ee
where we have used the first identity of eq.~(\ref{PPi}), and introduced the parameters\footnote{This determines $b_i$ up to an $i$-independent factor since ${\cal P}_i(z_i)$ is unambiguously defined, see eq.~(\ref{Pi}), and since ${\cal P}_i(z)$ is determined by the covering map constraint up to an overall constant.}
\be
(b_i)^{-1} = {\cal P}_i(z_i)\, {\cal P}^+(z_i) \ .
\ee
Since $m^i_2 \mapsto m^i_2+\frac{1}{2}$ maps $j_i\mapsto j_i + \frac{1}{2}$ and $u_i\mapsto u_i-\frac{1}{2}$, this implies that the constant $C(u_l;x_l,z_l)$ in eq.~(\ref{eq:SolutionC}) satisfies the recursion relation 
\be
C(u_i-\tfrac{1}{2}, u_j+\tfrac{1}{2};x_l,z_l) b_i a_i^{-\frac{1}{2}} = C(u_l;x_l,z_l) b_j a_j^{-\frac{1}{2}}  \ , 
\ee
and hence that the correlator is of the form 
\begin{align}
\langle B(z_5)B(z_6)&V^{w_1}_{m^1_1, m^1_2}(z_1,x_1)V^{w_2}_{m^2_1, m^2_2}(z_2,x_2)V^{w_3}_{m^3_1, m^3_2}(z_3,x_3)V^{w_4}_{m^4_1, m^4_2}(z_4,x_4)\rangle \nonumber \\
&\quad = C({\cal S};x_l,z_l)\, \prod_{i=1}^4 a_i^{ - 2 m_1^i}  b_i^{2 (m_1^i - m_2^i)}\ , \label{eq:SolutionC} 
\end{align}
where 
\be\label{Sdef}
{\cal S} = \sum_{i=1}^4 m_1^i-m_2^i=2+\sum_{i=1}^4 u_i \ . 
\ee
Furthermore, the $U_0$ charge conservation law requires that ${\cal S}=0$, and hence the above considerations determine the full $(m^i_1,m^i_2)$ dependence of the correlator. Thus the correlator is determined up to an unknown function $C({\cal S};x_l,z_l) \equiv \delta(x_4 - \Gamma(z_4))\delta_{{\cal S},0} C(x_l,z_l)$, where we have also restored the delta function dependence mentioned earlier.

\end{document}